\documentclass[11pt]{article}
\usepackage{epsf,epsfig,psfig,float,cite,amssymb,latexsym}
\newcommand{\La}{{\cal L}}  
\newcommand{\ci}{{\cal I}}
\newcommand{\cZ}{{\cal Z}}

\newcommand{\vp}{{\mathbf{p}}}
\newcommand{\vq}{{\mathbf{q}}}
\newcommand{\vx}{{\mathbf{x}}}
\newcommand{\vy}{{\mathbf{y}}}
\newcommand{\vQ}{{\mathbf{Q}}}
\newcommand{\vP}{{\mathbf{P}}}
\newcommand{\tQ}{{\widetilde{Q}}}

\newcommand{\Op}{{\mathcal{O}}(p)}  
\newcommand{\Opd}{{\mathcal{O}}(p^2)}  
\newcommand{\Opt}{{\mathcal{O}}(p^3)}  
\newcommand{\Opc}{{\mathcal{O}}(p^4)}  
\newcommand{\Opf}{{\mathcal{O}}(p^5)}  
\newcommand{\Ops}{{\mathcal{O}}(p^6)}

\newcommand{\barro}{\bar{\rho}}
\newcommand{\hatr}{\hat{\rho}}

\newcommand{\be}{\begin{equation}}  
\newcommand{\ee}{\end{equation}}  
\newcommand{\ba}{\begin{eqnarray}}  
\newcommand{\ea}{\end{eqnarray}}  
  
\newcommand{\nn}{\nonumber}  
  
\newcommand{\barr}[1]{\not\mathrel #1}  
  
\newcommand{\beq}{\begin{equation}}  
\newcommand{\eeq}{\end{equation}}  
\newcommand{\beqa}{\begin{eqnarray}}  
\newcommand{\eeqa}{\end{eqnarray}}  
  
\newcommand{\krig}[1]{\stackrel{\circ}{#1}}  
\newcommand{\vs}{\vspace{-0.20cm}}

\begin{document}

\hfill {\tiny FZJ-IKP(TH)-2001-14}

\thispagestyle{empty}

\vspace{2cm}

\begin{center}
{\Large{\bf In-medium chiral perturbation theory\\[0.3em]
beyond the mean--field approximation}}
\end{center}
\vspace{.5cm}

\begin{center}
{\large Ulf-G. Mei{\ss}ner\footnote{email: u.meissner@fz-juelich.de},  
  Jos\'e A. Oller\footnote{Present address: Universidad de Murcia, 
  Departamento de F\'{\i}sica, E-30071 Murcia, Spain \\
  email: oller@um.es},
  Andreas Wirzba\footnote{email: a.wirzba@fz-juelich.de}}
\end{center}

\begin{center}
{\it {\it Forschungzentrum J\"ulich, Institut f\"ur Kernphysik (Theorie) \\ 
D-52425 J\"ulich, Germany}}
\end{center}
\vspace{1cm}

\begin{abstract}
\noindent
{\small{An explicit expression for the generating functional of two--flavor
    low--energy QCD with external sources in the presence of non-vanishing
    nucleon densities has been derived recently~\cite{med1}.  Within this
    approach we derive power counting rules for the calculation of in-medium
    pion properties.  We develop the so-called standard rules for residual
    nucleon energies of the order of the pion mass and a modified scheme
    (non-standard counting) for vanishing residual nucleon energies. We also
    establish the different scales for the range of applicability of this
    perturbative expansion, which are $\sqrt{6}\pi f_\pi\simeq 0.7$ GeV for
    the standard and $6\pi^2 f_\pi^2/2m_N\simeq 0.27$ GeV for non-standard
    counting, respectively.  We have performed a systematic analysis of
    n--point in-medium Green functions up to and including next-to-leading
    order when the standard rules apply. These include the in-medium
    contributions to quark condensates, pion propagators, pion masses and
    couplings of the axial-vector, vector and pseudoscalar currents to pions.
    In particular, we find a mass shift for negatively charged pions in heavy
    nuclei, $\Delta M_{\pi^-}=(18\pm 5)\,{\rm MeV}$, 
    that agrees with recent determinations from deeply bound pionic
    $^{207}$Pb.  We have also established the absence of in-medium
    renormalization in the $\pi^0 \to \gamma\gamma$ decay amplitude up to the
    same order. The study of $\pi\pi$ scattering requires the use of the
    non-standard counting and the calculation is done at leading order. Even
    at that order we establish new contributions not considered so far.  We
    also point towards further possible improvements of this scheme and touch
    upon its relation to more conventional many-body approaches.  }}
\end{abstract}

\vspace{2cm}

\begin{center}
\end{center}

\bigskip

\centerline{Accepted for publication in {\it Ann. Phys.}}

\newpage

\section{Introduction}
\def\theequation{\arabic{section}.\arabic{equation}}
\setcounter{equation}{0}

The pion plays a special role in nuclear and particle physics. This is related
to the fact that for the light up and down quarks, QCD possesses an
approximate chiral symmetry, i.e. it is a good first approximation to the
theory to consider the light quarks as massless.  This symmetry is not present
in the ground-state or the particle spectrum; it is believed to be
spontaneously broken down to its vectorial subgroup, SU(2)$_L \times $SU(2)$_R
\to\,\,$SU(2)$_{L+R}$, with the appearance of three (Pseudo-)Goldstone bosons
which can be identified with the three pion states, $\pi^\pm, \pi^0$. The
chiral symmetry is also explicitly broken because the current quarks have a
small mass (small compared to a typical hadronic scale of 1~GeV). The unique
order parameter signaling this symmetry violation is the finiteness of the
weak pion decay constant in the chiral limit, denoted $f$, that is $f^2 \neq
0$. Another order parameter often considered is the quark condensate in the
vacuum, $\langle 0| \bar q q |0\rangle$, where $|0\rangle$ denotes the highly
complicated vacuum. However, it is important to stress that, in principle,
chiral symmetry could be broken even if $\langle 0| \bar q q |0\rangle \simeq
0$, as long as $f^2 \neq 0$. Due to Goldstone's theorem, the interactions of
the pions with themselves or matter fields must vanish as three-momentum and
energy go to zero. This in turn allows for a systematic treatment of such
processes in the framework of an effective field theory (chiral perturbation
theory, henceforth CHPT\cite{wph,gl}).  It is also believed that with
increasing temperature and/or density, the chiral symmetry of QCD is restored.
While lattice studies indicate that the critical temperature is $T_c \simeq
150\,$MeV, much less is known about the critical density. There have also been
recent speculations of a very complex phase structure at high densities (for a
recent review with many references, see \cite{rawi}). It is also important to
stress that lattice QCD applied to finite chemical potential $\mu$ (density
$\rho$) is only in its infancy due to the notorious sign problem of the
Euclidean Dirac operator at $\mu \neq 0$ (for a recent method to tackle this
problem see \cite{fodor} and references therein).  This provides another
reason why it is necessary to develop an in-medium effective field theory.
Pion properties, calculated at finite temperature and/or density, are
therefore used to gain an understanding how these transitions are approached
and in general terms to improve our knowledge of QCD at finite density.
Another important development are the recent measurements of deeply bound
pionic states in heavy nuclei performed at GSI \cite{g,i}. These experimental
data can be interpreted in terms of a pion mass shift due to the high nuclear
density in the center of such nuclei, and these have triggered some
calculations making use of heavy baryon chiral perturbation theory to unravel
the physics behind this intriguing phenomenon, see e.g. refs.~\cite{ko,kw1}.

In this paper we will undertake a {\em systematic} study of the properties of
pions and external sources in nuclear matter. Nuclear matter is a system of an
infinite number of protons and neutrons. For typical densities, these
interactions of the constituents of nuclear matter are strong, that is one
deals with phenomena in the non-perturbative regime of QCD. Therefore,
approximations are unavoidable, but one needs to be able to control these, as
it is possible using effective field theory.  Consequently, the final aim of
an in-medium QCD effective field theory is to provide a systematic way of
proceeding which allows one to estimate the errors when truncating the
expansion at some order.  The low energy effective field theory of QCD is
CHPT. Indeed CHPT allows not only to tackle processes involving pions but as
well to consider nucleons (baryons). These massive states are included as
matter fields chirally coupled to pions and external sources.  As stated, CHPT
in the vacuum is a systematic way of proceeding. Thus there are many articles
in the literature \cite{kaplan,other,aw,aw2} which apply CHPT Lagrangians that
are at most bilinear in the nucleon fields to the nuclear case in the
following way: The bilinears $\bar{N}\,D\,N$ (with $D$ a generic differential
operator including the coupling to pions and external fields) are replaced, in
the non-relativistic mean-field approach, by $\rho_p {\rm tr} D_{11}+\rho_n
{\rm tr} D_{22}$. Here $\rho_p \, (\rho_n)$ are the proton (neutron) density,
the symbol ${\rm tr}$ refers to the trace over spinor indices and the
subscripts run in flavor space.  Proceeding in this way one keeps track of the
information contained in the vacuum CHPT Lagrangians, but the chiral counting
in the medium is lost since nucleon correlations, due to the baryon
propagators, are not considered. In fact, such contributions can be of the
same or even of lower chiral order as those terms kept in the mean-field
approach. As shown below, they are the dominant contributions when the energy
flowing through the baryon propagator is of the order of a nucleonic kinetic
energy. However, as in the mean-field approach, we will not consider
multi-nucleon local interactions in this paper.  For attempts to develop 
effective field theories in nuclear matter without pions, see, e.g.,
refs.~\cite{fur,steele} and references therein. It is also important to stress
that our approach does not only encompass, but also exceeds the so--called
low--density theorems as formulated in refs.~\cite{LDT}.

In order to go beyond the mean field approach we follow ref.~\cite{gl} and
consider the CHPT Lagrangian supplemented by external fields. We also use the
results from ref.~\cite{med1} where the in-medium contribution to the
SU(2)$\times$SU(2) generating functional is calculated. Consequently, the
in-medium CHPT Lagrangian, in terms of {\it only} pions and external sources,
is given. At this stage, the problem reduces to that of vacuum CHPT, except
for the important difference that the resulting Lagrangian is non-covariant as
well as non-local (for a general analysis of the structure of non-relativistic
local effective field theories see \cite{leut}).  These findings are reviewed
in section \ref{sec:gf} and appendix \ref{subsec:genfunc} where the in-medium
chiral Lagrangian is given in terms of the vacuum CHPT Lagrangian and the
proton and neutron densities.  The chiral expansion is discussed in appendix
\ref{sec:operators}, while the (chiral) power counting is established in
sections \ref{sec:counting} and \ref{sec:counting-b}.  We then apply this
machinery in sections \ref{sec:qkcon}$-$\ref{sec:psc} to evaluate the
in-medium quark condensates, pion propagation and masses, pion couplings to
the axial-vector, vector and pseudoscalar currents and $\pi\pi$ scattering. In
addition, the decay $\pi^0\rightarrow \gamma\gamma$ is also studied and the
relevant scales of the expansion are discussed. We end with some conclusions
in section \ref{sec:conclusions}. Various technical topics are relegated to
the appendices.

\section{Generating functional, effective Lagrangian and chiral counting}  
\label{sec:gf}
\def\theequation{\arabic{section}.\arabic{equation}}
\setcounter{equation}{0}

For completeness, we briefly review in appendix \ref{subsec:genfunc} the main
result of ref.~\cite{med1}, repeatedly used in this work, where the in-medium
chiral effective Lagrangian $\widetilde{\La}_{\pi\pi}$ is derived by
integrating out the nucleon fields with functional methods. According to
Ref.\,\cite{wein}, a general chiral Lagrangian can be expanded in an
increasing number of baryon fields, $\La=\La_{\pi\pi}+\La_{\bar{\psi}\psi}+
\La_{\bar{\psi}\bar{\psi}\psi \psi}+...$, where $\psi$ denotes here the
nucleon field. The results of ref.~\cite{med1} were obtained by truncating
this series at terms bilinear in the spinor fields, thus neglecting the
multi-nucleon Lagrangians with four or more $\psi$ fields. The reason for this
is twofold. First, one does not know how to perform quartic Feynman path
integrals exactly.  Thus, the analysis of ref.\cite{med1} cannot be extended
in a straightforward way to include multi-nucleon interactions beyond a pure
perturbative treatment of those terms already discussed in ref.\cite{med1}.
Second, and related to the latter point, vacuum multi-nucleon interactions are
{\it nonperturbative} due to the extremely large $NN$ scattering lengths and
hence some kind of resummation is needed to end with an in-medium effective
field theory of nuclear matter which also includes multi-nucleon local
interactions together with pions.  Nevertheless, the situation is encouraging
due to the advances in understanding the multi-nucleon interactions in vacuum
by the application of effective field theory to such systems. Nowadays one has
two effective field theory schemes to tackle such problem, the original
Weinberg scheme \cite{wein} (which has been made truly quantitative in
\cite{EGM}) and the Kaplan-Savage-Weise one \cite{KSW}. We also refer to a
recent and comprehensive review on this topic\cite{laconga}.  How these
advances can be applied to the nuclear medium leading to a satisfactory
effective field theory is beyond present knowledge, see, e.g. ref.~\cite{fur}
for further discussions on this issue where the idealized case of natural
interactions without pions is considered in a Fermi system. Nevertheless, we
view it as a step forward to extend the effective field theory techniques used
in vacuum CHPT\cite{gl} to the medium in the case where pions and external
sources are kept, but the multi-nucleon local interactions are neglected. This
was done in ref.\cite{med1} and here we will exploit this formalism by
calculating systematically several in-medium Green functions within this
framework.

In appendix \ref{sec:operators} the expressions of the first and second order
interaction operators $A^{(1)}$ and $A^{(2)}$, as defined in appendix
\ref{subsec:genfunc} following ref.~\cite{med1}, are obtained from $\La_{\pi
  N}$.  The operator $A$ is given by the difference between the full and the
free Dirac operator (i.e.\ $D$ and $D_0$, respectively, see eq.\,(\ref{defa}))
and is amenable to a systematic chiral expansion as indicated by the
superscripts ``(1,2)'', see eqs.\,(\ref{a1}) and (\ref{pin2}).

In secs.~\ref{sec:counting} and \ref{sec:counting-b} we establish the chiral
counting of a given diagram resulting from $\widetilde{\La}_{\pi\pi}$, such
that we can determine the contributions up to a given order. We point out that
one has two different counting schemes depending on the energy flow through
the baryon propagators.


\subsection{Chiral counting: Standard case}
\label{sec:counting}  

\begin{figure}[htb]
\centerline{\epsfig{file=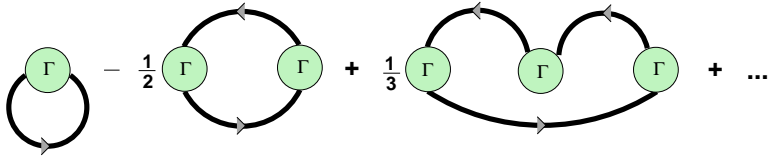,width=4.0in}}
\vspace{0.3cm}
\caption[pilf]{\protect \small
  Diagrammatic expansion of eq.\,(\ref{fZ2}). Every thick line corresponds to
  the insertion of a Fermi-sea and each circle to the insertion of an operator
  $\Gamma\equiv -i A\left[I_4-D_0^{-1}A\right]^{-1}$. Notice that each of the
  three in-medium baryon closed loops shown in the figure is a different
  generalized in-medium vertex.
\label{fig:gv}}
\end{figure}

In this section, we establish the power counting rules of the
processes that result from the in--medium generating functional derived 
in \cite{med1} and discussed in appendix \ref{subsec:genfunc}. Our starting 
point is the general structure of an in-medium generalized
non-local vertex, see fig.~\ref{fig:gv}.

\begin{figure}[htb]
\centerline{\epsfig{file=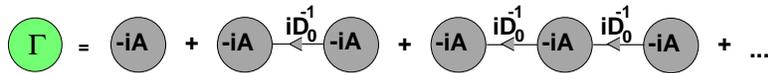,width=4.0in}}
\vspace{0.3cm}
\caption[pilf]{\protect \small
  Expansion of the non-local vacuum vertex $\Gamma\equiv=-i A\left[ I_4
    -D_0^{-1} A\right]^{-1}$, where $I_4$ is the $4\times 4$ unit matrix.
  Every solid line corresponds to a vacuum baryon propagator $iD_0^{-1}$ and
  each circle to the insertion of an operator $-i A$ from $\bar{\psi} D\psi$.
\label{fig:gop}}
\end{figure}

\noindent
As sketched in appendix~\ref{subsec:genfunc} and detailed in \cite{med1}, each
thick solid line corresponds to summing over all the plane wave states of the
proton(neutron) Fermi-sea with three-momentum smaller than
$k_F^{(p)}$($k_F^{(n)}$).  In the following we count any Fermi momentum $k_F$
(about $\sim 2 M_\pi$ for nuclear saturation density $\rho_0=0.17$ fm$^{-3}$)
as $M_\pi \sim \Op$. Then each thick line, because of the three-momentum
integration up to the Fermi momentum, is $\Opt$. Next, consider the non-local
vacuum vertices $\Gamma\equiv -iA[I_4-D_0^{-1}A]^{-1}$, see
fig.\,\ref{fig:gop}.

\noindent
Given the pion-nucleon Lagrangian ${\cal L}_{\pi N}=\bar{\psi}D(x)\psi$, the
operator $A$ is itself subject to a chiral expansion, $A=\sum_i A^{(i)}$,
$i\geq 1$ (the terms in $A$ up to $\Opd$ are explicitly worked out in
appendix~\ref{sec:operators}).  On the other hand, a soft momentum $Q$ related
to pions or external sources can be attached to each local vertex $A$. This
together with the Dirac delta function of four-momentum conservation implies
that the momenta running along the baryon propagators $D_0^{-1}$ just differ
from each other by quantities of order $Q\sim \Op$. Since at least one
Fermi-sea insertion (thick line) is required for any in-medium contribution,
with on-shell four-momentum $p$\footnote{Note that we use the same symbol $p$
to denote an on-shell Fermi-sea four-momentum, which is ${\cal O}(1)$, or to
indicate the chiral dimension of some contribution as ${\cal O}(p^n)$ and in
this case $p\sim {\cal O}(p)$.  Nevertheless, for later use, $p$ is always
accompanied by the symbol ${\cal O}$ or the word {\it order} and no
confusion should appear.}, the four-momentum $p_j$ running along the
$j^{th}$ free baryon propagator $iD_0^{-1}(p_j)$, is $p_j=p+Q_j$. In this way
the natural chiral power counting of such propagators can be obtained by
expanding them as,
\ba
 \label{pop}
 iD_0^{-1}(p_j)&=&i\frac{\barr{p}+\barr{Q}_j+m_N}{(p+Q_j)^2-m_N^2+i\epsilon}=
 i \frac{\barr{p}+\barr{Q}_j+m_N}{Q_j^2+2 Q_j^0 E_N(p) -2 {\mathbf{Q}}_j 
 \mathbf{p}+i\epsilon}
 \nonumber \\
 &=&i \frac{\barr{p}+\barr{Q}_j+m_N}{2 Q^0_j m_N+i\epsilon} 
 \left(1-\frac{Q_j^2-2 {\mathbf{Q}}_j \mathbf{p}}{2Q_j^0 m_N}+\Opd \right)~.
\ea 
Thus $iD_0^{-1}$ counts as ${\cal O}(p^{-1})$. As a result every $\Gamma$
vertex with $m_\Gamma\geq 0$ free baryon propagators and $(m_\Gamma+1)$ $A$
vertices (see fig.\,\ref{fig:gop}) scales as $p^{\nu_\Gamma}$ with
\ba 
 \label{cg}
 \nu_\Gamma=\sum_{j=1}^{m_\Gamma+1}d_j-m_\Gamma~,
\ea
where $d_j$ is the chiral dimension of the vertex $A_j$. Next, we will
consider the chiral counting of {\em in-medium non-local} vertices, some of
them are shown in fig.\,\ref{fig:gv}.  The chiral counting of an in-medium
non-local vertex (labeled by $\rho$) that is generated through the exchange of
$n\geq 1$ Fermi-sea insertions, $n$ non-local vacuum $\Gamma$ vertices,
$V=n+m$ local $A$ vertices and $m=\sum_i m_{\Gamma_i}$ free baryon propagators
is given by 
\be
 \label{c2}
 d_\rho=3n+\sum_{i=1}^n \nu_{\Gamma_i}-4(n-1)=3n+\sum_{i=1}^V d_i-m-4(n-1)=
 \sum_{i=1}^V d_i-V+4=4+\sum_{i=1}^{V}(d_i-1)~.
\ee
Here the factor $4(n-1)$ originates from the four-momentum Dirac deltas
attached to any $\Gamma$ vertex. Now, since $d_i\geq 1$, any such diagram will
count at least as ${\cal O}(p^4)$. The lowest order is obtained when only
$A^{(1)}$ operators with $d_i=1$ are included. The next-to-leading order
implies the inclusion of one $A^{(2)}$ operator and is $\Opf$.\footnote{In the
  following whenever we say that some calculation is done up to order $p^m$,
  ${\cal{O}}(p^m)$, we mean up--to--and--including contributions of order
  $p^m$. It is also important to stress that the order we give always refers
  to the order of the terms in the Lagrangian (generating functional) that
  have been used to calculate a particular quantity. Because observables are
  obtained by differentiation of the generating functional with respect to the
  external sources, the leading and subsequent higher orders might be reduced
  by a {\it common} (or more) power(s) in $p$. This should be kept in mind
  throughout.}

Finally one has to take into account the exchange of pions inside or between
generalized in-medium vertices or between them and pure pion vertices coming
from the vacuum chiral Lagrangian ${\cal L}_{\pi\pi}$ or between the latter
alone.  Denoting by $I_\pi$ the total number of internal pion lines and by
$L_\pi$ the numbers of pion loops, $L_\pi=I_\pi-V_\rho-V_\pi+1$ where $V_\pi$
is the number of vertices from ${\cal L}_{\pi\pi}$ and analogously $V_\rho$ is
the number of generalized in-medium vertices. Hence a many-particle diagram
originating from the Lagrangians ${\cal L}_{\pi\pi}$ and ${\cal L}_{\pi N}$
has the chiral power $\nu$,
\ba
 \label{fc}
 \nu
 =4L_\pi-2I_\pi+\sum_{i=1}^{V_\pi}d_i+\sum_{i=1}^{V_\rho}d_{\rho i}=2L_\pi+2+
 \sum_{i=1}^{V_\pi}(d_i-2)+\sum_{i=1}^{V_\rho}(d_{\rho i}-2)=2L_\pi+2+
 \sum_{i=1}^{V_T}(\delta_i-2)
\ea 
with $V_T=V_\pi+V_\rho$ and $\delta_i$ is the chiral dimension of any vertex,
either from ${\cal L}_{\pi\pi}$ or from the in-medium generalized vertices.
We do not differentiate between $d_i$ and $d_{\rho i}$ and use the symbol
$\delta_i$ wherever no confusion can arise whether a vertex comes from the
free--space Lagrangian or is an in-medium generalized one.  The counting based
on eq.\,(\ref{fc}) is from here on referred to as the {\em standard case}.

As a result the leading in-medium contribution begins at ${\cal O}(p^4)$,
since $d_\rho \geq 4$ from eq.\,(\ref{c2}) with $V_\rho=1$ and $L_\pi=0$. In
this work we will calculate the next-to-leading order in-medium corrections,
${\cal O}(p^5)$, to several $n-$point Green functions whenever the previous
counting holds, see below.  The $\Opc$ vacuum corrections to the leading
non-vanishing $\Opd$ results can be found in ref.~\cite{gl}, except for the
anomalous $\pi^0\to \gamma\gamma$ decay (which only starts at fourth order in
the chiral expansion since it is an anomalous process), and when necessary we
will just give the corresponding vacuum results. As it is clear from
eq.\,(\ref{fc}), any pion loop or any extra generalized in-medium
vertex\footnote{ Henceforth we will independently use the words in-medium or
  the symbol $\rho-$ to denote any contribution due to the finite nucleon
  density of the ground state. The symbol $\rho-$ should not be confused with
  the $\rho$ meson, which we always denote as $\rho (770)$.} increases the
chiral power by at least two more orders, ${\cal O}(p^6)$. Thus the leading
$\rho-$corrections, according to eqs.\,(\ref{c2}), (\ref{fc}) are of order
$p^4$ and $p^5$ and come from the insertion of one generalized $\rho-$vertex
with only $A^{(1)}$ operators or with one additional $A^{(2)}$, respectively.
We state once again that it is beyond the scope of the present work, and
beyond the present knowledge, to derive a full in-medium effective field
theory which includes as well nucleon-nucleon contact interactions.


\subsection{Chiral counting: Non-standard case}
\label{sec:counting-b}  

There is one subtlety with the power counting developed so far. To understand
this, we remark that the propagator of a nucleon in the heavy baryon
formulation with four-momentum $p_j$ is $1/(Q_j^0+i\epsilon)$. This
corresponds to the leading non-relativistic term of the expansion given in
eq.\,(\ref{pop}) which was used in order to determine the dimension of
$iD_0^{-1}(p_j)$.  One caveat, already present in vacuum, arises when the
energy $Q_j^0$ is fixed and vanishes\footnote{The expansion of
  eq.\,(\ref{pop}) is of quantities $Q^2\sim M_\pi^2\sim k_F^2$ over $2Q_j^0
  m_N$ and then, when $Q_j^0$ is of the order $k_F^2/2m_N$, it breaks down.}.
In this case the heavy baryon propagator blows up although in a relativistic
formalism used here (or in a non-relativistic one by keeping the kinetic
energy) one still has
$iD_0^{-1}(p_j)=-i(\barr{p}+\barr{Q}_j+m)/({\mathbf{Q}}_j^2+2{\mathbf{Q}}_j
{\mathbf{p}}+i\epsilon)$.  The result is finite, but now $i D_0^{-1}(p_j)$
counts as ${\cal O}(p^{-2})$ instead of ${\cal O}(p^{-1})$ as it was used to
derive eq.\,(\ref{fc}). To deal with this case, it is necessary to subtract
the number of baryon lines, $I_B^\star$, with $Q_j^0\lesssim k_F^2/2m_N$ from
eq.\,(\ref{fc}).  We can write an upper bound of this number in terms of the
number of loops, vertices and external lines.  To see this note that the
number of pion plus external source legs is greater or equal to the number of
baryon lines, which can be free propagators as well as Fermi-sea insertions.
We make use of the fact that each $A$ vertex will involve at least one pion or
external source, otherwise it is absorbed in the physical value of the nucleon
masses, as explained in appendix~\ref{sec:operators}.  We denote by $E_x$ the
number of external legs and by $I_B$ the number of baryon lines {\em minus}
$V_\rho$, since for every generalized vertex we have at least one Fermi-sea
insertion. Then we have 
 \be 
 E_x+2I_\pi-V_\rho \geq I_B~.  
\ee Substituting in
this expression $I_\pi=L_\pi+V_T-1$, one obtains 
\be
 \label{nc2}
 E_x+2L_\pi+V_\rho+2V_\pi-2\geq I_B~.
\ee  
We can further restrict the previous inequality by subtracting $2V_\pi$ on the
left hand side (l.h.s.) since for each vacuum vertex there are at least two
legs attached to it when calculating any in-medium contribution. Moreover, in
the case of diagrams without any external source we can subtract $4V_\pi$
because any vacuum vertex should have at least four pion legs. Otherwise it
just contributes to the vacuum renormalized pion propagators and associated
wave function renormalizations, which can be calculated anyhow just from
${\cal L}_{\pi\pi}$ to the required accuracy.  We will denote by a) the first
case and by b) the more specific second one. Then we can write:
\ba
 \label{nc3}
 &\hbox{a) }& E_x+2L_\pi+V_\rho-2 \geq I_B~, \nn \\
 &\hbox{b) }& E_x+2L_\pi+V_\rho-2-2 V_\pi \geq I_B~.
\ea
We now indicate by $n_m$ the number of $A$ vertices with $m$ attached legs.
Then for each of such vertices $m-1$ lines are non-baryonic and can be removed
from the l.h.s of eqs.\,(\ref{nc3}). In the same way, we call $p_m$ the number
of vacuum vertices from ${\cal L}_{\pi\pi}$ with $m$ legs. For the case a) we
have already subtracted $-2V_\pi$ and for the case b) $-4V_\pi$. Thus we can
further remove $m-2$ for a) and $m-4$ for b). Putting all this together, we
arrive at the equalities:
\ba
 \label{nc4}
 &\hbox{a) }& 
 E_x+2L_\pi+V_\rho-2-\sum_{m\geq2}(m-1)n_m-\sum_{m\geq3}(m-2)p_m=I_B~, \nn \\
 &\hbox{b) }&
  E_x+2L_\pi+V_\rho-2-2V_\pi-\sum_{m\geq2}(m-1)n_m-\sum_{m\geq5}(m-4)p_m=I_B~.
\ea
Notice that the second of these equations is just a rewriting of the first one
for the special case with only pion legs. In summary, after the elimination of
$L_\pi$ and the subtraction of $I_B^\star$, the counting index $\nu$ given in
eq.\,(\ref{fc}) changes to
\ba
 \label{ffc}
 &\hbox{a) }&
 \nu=4+\sum^{V_\rho}_{i=1}(\delta_i-3)+\sum^{V_\pi}_{i=1}(\delta_i-2)-E_x+
 \sum_{m\geq2}(m-1)n_m+\sum_{m\geq3}(m-2)p_m+\left(I_B-I_B^\star\right)~,
 \nn\\
 &\hbox{b) }&
 \nu=4+\sum^{V_\rho}_{i=1}(\delta_i-3)+\sum^{V_\pi}_{i=1}\delta_i-E_x
 +\sum_{m\geq2}(m-1)n_m+\sum_{m\geq5}(m-4)p_m+\left(I_B-I_B^\star\right)~. 
\ea
When $V_\pi=0$, we adopt the convention that there is no contribution from the
second and fourth sums on the right hand side (r.h.s) of the last equation.
The interesting point of these equations is that for a given number of
external legs $E_x$ the chiral dimension $\nu$ is bounded since $\delta_i \geq
4$ for a generalized in-medium vertex as shown above and $\delta_i\geq 2$ for
any vertex from ${\cal L}_{\pi\pi}$ and by definition $I_B\geq I_B^\star$.
Note also that the inclusion of an extra generalized $\rho-$vertex increases
$\nu$ at least by 1. Those loops that arise because of pion lines that are
exchanged inside only one generalized $\rho-$vertex will increase
$I_B-I_B^\star$ by the number of involved baryon propagators in the loops
because of the absence of the so called pinch singularity~\cite{wein}.
However, when two or more $\rho-$vertices are involved in the pion loops this
is no longer the case. We will apply these counting rules when discussing
in-medium pion condensation and $4\pi$ scattering. In the latter case, we will
just calculate the leading contribution $\Opt$. The counting based on
eq.\,(\ref{ffc}) is from here on referred to as the {\em non-standard case}.

In the sections \ref{sec:qkcon}--\ref{sec:anom}, we will use the simpler
counting rules, see eq.\,(\ref{fc}), to determine the diagrams to be
considered up to $\Opf$, with the exception of those cases cases where a small
energy denominator occurs, $Q_j^0\simeq k_F^2/2m_N$, and where we will turn to
eq.\,(\ref{ffc}). This will always be indicated in the text at the appropriate
places.

Finally, let us stress again the difference in the power counting for pion
properties in the medium and in the vacuum. While in vacuum the leading and
next-to-leading order is ${\cal O}(p^2)$ and ${\cal O}(p^4)$, respectively, in
the medium and for the standard case, these contributions start at 
${\cal  O}(p^4)$ and ${\cal O}(p^5)$, respectively.


\section{In-medium quark condensates}
\label{sec:qkcon}
\begin{figure}[htb]
\centerline{\epsfig{file=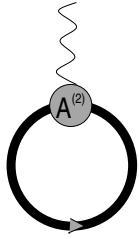,width=.7in}}
\vspace{0.3cm}
\caption[pilf]{\protect \small 
  Unique diagram that contributes to the calculation of the in-medium quark
  condensates up to $\Opf$. As usual the thick line represents a Fermi-sea
  insertion, the circle is an $A^{(2)}$ vertex and the wavy line an external
  scalar source $s(x)$.
\label{fig:tad}}
\end{figure}

\def\theequation{\arabic{section}.\arabic{equation}}
\setcounter{equation}{0}
{}From the generating functional ${\cal Z}(v,a,s,p)$ 
given in eq.\,(\ref{fZ2}), the quark condensate is obtained by  partial 
functional differentiation,
\be
 \langle \Omega| \bar{q}_i q_j|\Omega \rangle=-
 \frac{\delta}{\delta s_{ij}(x)}{\cal  Z}(v,a,s,p)\arrowvert_{v,a,s,p=0}~,
\ee
with $q_1=u$ and $q_2=d$ quarks. Up to ${\cal O}(p^5)$, because of the absence
of pion loops, the in-medium contributions can be calculated directly by
equating the generating functional ${\cal Z}$ with the in-medium action $\int
dx \widetilde{\cal L}_{\pi\pi}(x)$ and the vertex can be identified by doing
the partial derivative $\frac{\partial}{\partial s_{ij}(x)} A^{(2)}(x)$. The
resulting $\rho-$diagram is depicted in fig.\,\ref{fig:tad} with one insertion
from $A^{(2)}$, since the scalar source $s(x)$ is not present in $A^{(1)}$
(see eqs.\,(\ref{a1}) and (\ref{pin2})).

We denote by $\rho_p$ and $\rho_n$ the proton and neutron densities,
in order. These are given by 
\ba
 \label{den}
 \rho_p = 2\int^{k_F^{(p)}}\!\!\!\frac{d\mathbf{p}}{(2\pi)^3}=
 \frac{(k_F^{(p)})^3}{3\pi^2}~, \qquad
 \rho_n = 2\int^{k_F^{(n)}}\!\!\!\frac{d\mathbf{p}}{(2\pi)^3}=
 \frac{(k_F^{(n)})^3}{3\pi^2}~,
\ea
and we also introduce the quantities $\hat{\rho}=(\rho_p+\rho_n)/2$, the
isospin symmetric nuclear density, and $\bar{\rho}=(\rho_p-\rho_n)/2$, the
isospin asymmetric nuclear density. It is easy to see from eqs.\,(\ref{fZ2})
and (\ref{pin2}) that the in-medium contributions to the action $\int
dx\,\widetilde{\La}$ which contain the scalar sources (the ones relevant for
the calculation of the quark condensates) are:
\ba
 \label{qcact}
&&
 \int dx\int\frac{d\vp}{(2\pi)^3 2E(p)}\hbox{Tr}\Bigl[\left\{c_1 4B_0
 \langle s(x)\rangle+
 c_5\left[4B_0s(x)-2B_0\langle s(x) \rangle \right]\right\}n(p)
 (\barr{p}+m_N)\Bigr]\nonumber \\
 &&
 =\int dx\Bigl[4 B_0\hat{\rho} (2c_1-c_5)(s_{11}(x)+s_{22}(x))
 +4B_0 c_5(s_{11}(x)\rho_p+ s_{22}(x)\rho_n)\Bigr]+{\mathcal O}(p^6)~,
\ea
where the usual SU(2) matrix notation is applied for the scalar source.
By differentiating with respect to $s_{11}(x)$ and $s_{22}(x)$  one obtains, 
respectively, the up- and the down-quark condensates at $\Opf$:
\ba
 \label{qc}
 \langle \Omega|\bar{u}u|\Omega\rangle&=&\langle \bar{u}u \rangle_{\rm vac}
 \left[1-\frac{2\sigma}{f^2 M_\pi^2}\hat{\rho}
 +\frac{4 c_5}{f^2}\bar{\rho}\right]\nn~,\\
 \langle \Omega|\bar{d}d|\Omega\rangle&=&\langle \bar{d}d \rangle_{\rm vac}
 \left[1-\frac{2\sigma}{f^2 M_\pi^2}\hatr-\frac{4 c_5}{f^2}\barro\right]~,
\ea
with the pion-nucleon sigma-term $\sigma=-4c_1 M_\pi^2$ at
$\Opd$~\cite{gass,bkm1}. Furthermore, the subscript $\rm vac$ refers to the
vacuum value of the corresponding quantity up to $\Opc$ as given in
ref.~\cite{gl}, which at lowest order is simply $-B_0 f^2$.

Keeping terms up to the same accuracy,\footnote{In the following, when the
  previous sentence is not explicitly stated, the context will clarify when an
  equality is exact or accurate up to higher order corrections.}  the
isoscalar sum $\langle \Omega|\bar{u}u
+\bar{d}d|\Omega\rangle=\langle\bar{u}u+\bar{d}d\rangle_{\rm
  vac}\left[1-2\sigma \hatr/f^2M_\pi^2 \right]$ only depends on the total
nuclear density $2\hatr=\rho_p+\rho_n$ and agrees with the expression, valid
up to linear order in the density, given in ref.~\cite{furn} under the use of
the Hellmann-Feynman theorem (we are using a chiral perturbative approach and
the agreement is only valid strictly when calculating the $\sigma$-term up to
$\Opd$ as noted above). For a more intuitive derivation, see
ref.~\cite{levin}. This coincidence is not surprising since, in the
non-relativistic limit, the diagram of fig.\,\ref{fig:tad} just counts the
number of nucleon states times the corresponding vertex for protons and
neutrons separately. If one assumes an expansion in density on intuitive
grounds this has to be the leading result.  This is indeed the argument used
in ref.~\cite{levin}. Note that deviations from the relativistic limit in
fig.\,\ref{fig:tad} increase the counting by two orders. As a new aspect with
respect to previous works \cite{levin,furn,aw} we also consider strong isospin
breaking represented by the term with $c_5$ in eq.\,(\ref{qc}).  Other works
that also consider the calculation of the quark condensates in symmetric
nuclear matter are refs.~\cite{aw,other} within a mean-field approach and
refs.~\cite{bm,hat,rein,ch} using the Nambu-Jona-Lasinio model. Note also that
the quark condensates are essential inputs of in-medium QCD sum rules
techniques \cite{fur2,levin}.

For the case of symmetric nuclear matter, all the previous investigations led
to dropping quark condensates with increasing density. Specifically, using the
value for the $\sigma$--term, $\sigma= 44\pm 8 \pm 7$ MeV \cite{gass2}, one
obtains a relative reduction of a $(35\pm 9)\%$ with respect to its vacuum
value for nuclear matter density. This reduction of the quark condensate with
density is traditionally considered as a clear indication of chiral symmetry
restoration in the nuclear medium~\cite{bm}. At this point it is worthwhile to
stress that the quark condensate is not a unique order parameter of chiral
symmetry breaking. Even if it vanishes, chiral symmetry is still spontaneously
broken as long as $f^2 \neq 0$. We will come back to this point below when
evaluating the in-medium weak decay couplings of the pion. We also note from
eq.\,(\ref{qc}) that the correction becomes unity when $\rho\sim 2.8 \rho_0$
corresponding to a Fermi-momentum $k_F\sim 400$ MeV. However, this value
should also be considered indicative since then the correction is equal to the
leading term.

Consider now the isospin breaking contribution.  {}From ref.~\cite{bkm1} we
take the value $c_5=-0.09\pm 0.01$ GeV$^{-1}$. For realistic cases $\rho_n\geq
\rho_p$ and then since $c_5<0$, the in-medium $\bar{d}d$ condensate is smaller
than the $\bar{u}u$ one.  Using $\langle \bar{u}u\rangle_{\rm vac}=\langle
\bar{d}d \rangle_{\rm vac}= \frac{1}{2}\langle\bar{q}q \rangle_{\rm vac}$ in
the limit $m_u=m_d$ \cite{gl}, the pertinent difference is: 
\be
 \label{dqc}
 \langle\Omega|\bar{u}u-\bar{d}d|\Omega\rangle=\langle\bar{q}q\rangle_{\rm vac}
 \frac{4\,c_5}{f^2}\barro\simeq\langle\bar{q}q\rangle_{\rm vac} \, \,
 0.028 \,
 \frac{\rho_n-\rho_p}{\rho_0},
\ee
which is very suppressed as compared to the density dependence of 
the sum given by 
\be
 \label{sqc}
 \langle\Omega|\bar{u}u+\bar{d}d|\Omega\rangle \simeq 
 \langle\bar{q}q\rangle_{\rm vac}\left[
  1-0.35\frac{\rho_n+\rho_p}{\rho_0}\right] ~,
\ee
because $-2c_5$, see. eq.\,(\ref{qc}), is more than one order of magnitude
smaller than $\sigma/M_\pi^2\sim 2.3$ GeV$^{-1}$. Hence the presence of the
terms in eq.\,(\ref{qc}) proportional to $c_5$ does not alter appreciably the
reduction of the quark condensates at finite density present in the symmetric
nuclear matter case studied in refs.~\cite{bm,furn,aw}.

\section{Two-point Green functions}
\label{sec:tpgf}
\noindent
\def\theequation{\arabic{section}.\arabic{equation}}
\setcounter{equation}{0}

We now consider the propagation of pions in the medium. We will first derive
their equations of motion from which one can read off the in-medium pion mass.
We will also discuss pion-condensation illustrating the use of
eq.\,(\ref{ffc}). In addition the new scales for the in-medium chiral
expansion are also pointed out. The $\rho-$wave function renormalization of
pions needed to satisfy the quantum canonical commutation relations is derived
in appendix \ref{subsec:wfr}.  Finally, we will also study the couplings of
pions to the axial-vector and pseudoscalar currents in the medium and their
relation as dictated by a QCD Ward identity.

\subsection{Equations of motion}
\label{sec:prop}

\begin{figure}[htb]
\centerline{\epsfig{file=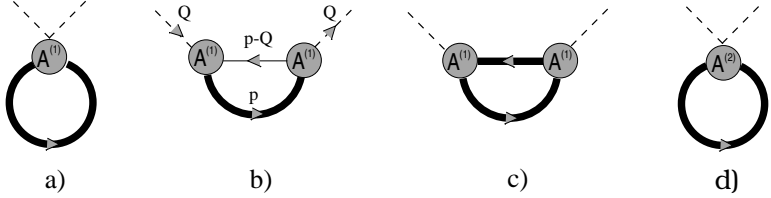,width=4.in}}
\vspace{0.3cm}
\caption[pilf]{\protect \small 
  Diagrams that contribute to the in-medium propagation of pions up to $\Opf$.
  Figs.a), b) and c) are $\Opc$ and fig.d) is $\Opf$. The crossed diagram of
  b), obtained by replacing $Q\rightarrow -Q$, is not shown. The energy
  through the pion lines (dashed) is $\sim M_\pi$.
\label{fig:pp}}
\end{figure}

In fig.\,\ref{fig:pp} we show those diagrams that give the in-medium
corrections of the pion propagator up to $\Opf$. As pointed out in
ref.~\cite{med1}, the $\rho-$vertices have analogous properties to those of
the standard local vertices (i.e.\ the field theoretical vertices from ${\cal
  L}_{\pi \pi}$) in the computation of the numerical factors accompanying the
exchange of pion lines in a given diagram. With respect to fig.~\ref{fig:pp}
this manifests itself in the fact that together with the momentum $Q$ flowing
from the left to the right, as in particular shown in fig.~\ref{fig:pp}b, one
has also to consider the inverse process, $Q\rightarrow -Q$.  This is
immediately accomplished when working in configuration space.

\noindent

Denoting by $Q$ the pion four-momentum and by $p$ the one running in the
(lower) Fermi-sea insertion ($p^2=m_N^2$) in each diagram of
fig.~\ref{fig:pp}, one can easily see that the imaginary parts of the diagrams
in this figure vanish for $Q^0\sim M_\pi \gg Q^2/(2 m_N)$. Note that the
presence of the pole of the baryon propagating with four-momentum $p-Q$ in
fig.\,\ref{fig:pp} implies the vanishing of the denominator
$(p-Q)^2-m_N^2=Q^2-2pQ$. This occurs only when $Q^0\simeq Q^2/2m_N$ which is
not the case for $Q^2\sim M^2_\pi$, i.e. for the type of pion momenta
considered in most parts of this work. The same reasoning can be applied to
fig.~\ref{fig:pp}c. Denoting by $q$ the four-momentum of the second Fermi-sea
insertion we then have $(p-Q)^2=q^2=m_N^2$ which is once again the same
equation as the one required to pick the pole in fig.\,\ref{fig:pp}b. In fact,
the imaginary parts of fig.~\ref{fig:pp}b and fig.~\ref{fig:pp}c are closely
linked and tend to cancel each other. Remembering the Feynman rules of
appendix~\ref{subsec:genfunc}, a second Fermi-sea insertion implies an extra
factor of 
\be
 \label{fse} 
 -\frac{1}{2}\int^{k_F}\!\!\frac{d{\mathbf q}}{(2\pi)^3 2 E(q)}(\barr{q}+m_N),
\ee 
as compared to the case with only one Fermi-sea insertion. On the other hand,
putting a baryon on-shell in fig.~\ref{fig:pp}b implies 
\be
 \label{bpos}
 i\int \frac{dq}{(2\pi)^4}\frac{\barr{q}+m_N}{q^2-m_N^2+i\epsilon}\rightarrow 
 \frac{1}{2}\int\frac{d\vq}{(2\pi)^3\, 2 E(q)}
 \left[(\barr{q}+m_N)-(\tilde{\barr{q}}-m_N)\right]~,
\ee
with the four-momentum $\tilde{q}=(E(q),-{\bf{q}})$. The second term on the
r.h.s. of this equation comes from the anti-baryon pole and does not
contribute because of the presence of an extra delta function
$\delta(q^0-p^0-Q^0_j)$ which cannot be satisfied since in this case
$q^0=-E(q)$ and $p^0$ is an on-shell baryon momentum coming from the sum over
the states in the Fermi-sea as already discussed.  Then only the first term on
the r.h.s. of eq.\,(\ref{bpos}) is possible and by splitting the
three-momentum integral as $\int^\infty d {\bf{q}}=\int^{k_F}d{\bf{q}}+
\int_{k_F}^\infty d{\bf{q}}$, we see that the integration up to $k_F$ cancels
against eq.\,(\ref{fse}), leaving only $\int_{k_F}^\infty d{\bf{q}}$.  This
implies that the absorptive parts, arising by exciting one baryon on-shell,
begin to contribute only when the baryon-three-momentum is larger than $k_F$
(particle excitation), as one could naturally expect since all the other
states in the Fermi-sea are already occupied.

Introducing the density matrix $\rho={\rm diag}(\rho_p,\rho_n)$, the in-medium
contributions to the quadratic pionic term of $\widetilde{{\cal L}}$, $\delta
\widetilde{{\cal L}}_{\phi\phi}$, can be obtained from eq.\,(\ref{fZ2}) (as
just discussed, the last term of eq.\,(\ref{fZ2}) with two Fermi-sea
insertions does not contribute here). We then obtain
\ba
 \label{pp1}
 i\int dx \delta \tilde{{\cal L}}_{\phi\phi}(x)
 &=&
 -\frac{1}{8f^2}\int dx \langle \rho [\phi,\dot{\phi}] 
 \rangle+\frac{i}{f^2}\int dx 
 \Big \langle \rho \bigg\{-2 c_1 B_0 \langle \phi^2 {\cal M} \rangle
 +\frac{1}{2}c_2\langle \dot{\phi} \dot{\phi}\rangle
 +\frac{1}{2}c_3\langle \partial_\mu \phi \partial^\mu \phi \rangle 
 \nn \\
 &-&
 \frac{1}{2}c_5 B_0 \left(\phi^2 {\cal M}+{\cal M} \phi^2+2\phi {\cal M} 
 \phi \right)+
 c_5 B_0\langle \phi^2 {\cal M}\rangle \bigg\}\Big\rangle \\
 &-&
 \frac{i g_A^2}{4 f^2}\int \frac{d\mathbf{p}}{(2\pi)^3 2E(p)}
 \int dx dy\, e^{ip(x-y)}
 \hbox{Tr}\bigg\{\gamma^\mu \gamma_5  \partial_\mu\phi(x) D_0^{-1}(x,y) 
 \gamma^\nu\gamma_5 \partial_\nu\phi(y)(\barr{p}+m_N)n(p)\bigg\},\nn 
\ea
where the Cartesian coordinates for the pion fields are defined by
$\phi=\sum_{i=1}^3\phi_i \tau_i$ with $\phi$ given in eq.\,(\ref{phi}) and
$\tau_i$ are the usual Pauli matrices. In the last expression we have set
$\krig{g}_A$ to its physical value $g_A=1.267$ since corrections will be of
higher orders than the next-to-leading one, $g_A=\krig{g}_A\{1+{\cal
  O}(\hat{m})\}$\cite{revij}, and for the same reason we take in the following
$m_p=m_n=m_N=(m_n+m_p)/2\simeq939\hbox{ MeV}$ as mentioned in
appendix~\ref{subsec:genfunc}.

The in-medium contribution to the equations of motion are obtained by
performing the differentiation $\delta \int dy
\widetilde{\La}_{\phi\phi}/\delta \phi_k(x)$.  Working in momentum-space,
$\phi_k(x)=\int \frac{dQ}{(2\pi)^4}\,e^{-iQx} \phi_k(Q)$, one obtains: 
\ba
 \label{eqm1}
 &-&
 \frac{g_A^2\hat{\rho}}{2f^2 m_N}\frac{(Q^2)^2}{Q_0^2}\phi(Q)_k-i
 \frac{g_A^2\bar{\rho}}{f^2}\left(Q_0-\frac{Q^2}{Q_0}\right)
 \epsilon_{kj3}\phi(Q)_j
 +i\frac{\bar{\rho}}{f^2}Q_0\epsilon_{kj3}\phi(Q)_j
 -\frac{16 B_0}{f^2}\hat{\rho}\hat{m} c_1 \phi(Q)_k
 \nn \\
 &+&
 \frac{4\hat{\rho}}{f^2}
 \left(c_2 Q_0^2+c_3 Q^2\right)\phi(Q)_k
 -\frac{8B_0}{f^2}\bar{\rho}\bar{m}c_5 
 \delta_{3k}\phi(Q)_3 ~.
\ea
Diagonalizing these equations by working in terms of the charged fields
$\pi^+=(\phi_1-i\phi_2)/\sqrt{2}$ and $\pi^-=(\phi_1+i\phi_2)/\sqrt{2}$,
together with $\pi^0=\phi_3$ and adding the vacuum piece, we finally have the
following spectral relations between the energy $Q_0\equiv \omega$ and the
three-momentum $\mathbf{Q}$ for on-shell in-medium pions:

\ba
 \label{spec}
 \mathbf{\pi^0:} && \nn\\
&&
 \omega^2-M_{\pi^0}^2\left(1+c_1\frac{8\hat{\rho}}{f^2}\right)
 +\frac{4\hat{\rho}}{f^2}\omega^2 
 \left(c_2+c_3-\frac{g_A^2}{8m_N}\right)
 -{\mathbf{Q}^2}\left(1+\frac{4\hat{\rho}}{f^2}c_3
 -\frac{g_A^2\hat{\rho}}{m_N f^2}\right)-\frac{g_A^2 \hat{\rho}}{2f^2 m_N}
 \frac{({\mathbf{Q}}^2)^2}{\omega^2} \nn \\
&&
 -\hat{M}_\pi^2 \frac{4\bar{\rho}}{f^2}
 \frac{\bar{m}}{\hat{m}}c_5=0~,
 \nn\\
 \mathbf{\pi^+:} 
&&\nn\\
&&
 \omega^2-M_{\pi^+}^2\left(1+c_1\frac{8\hat{\rho}}{f^2}\right)
 +\frac{4\hat{\rho}}{f^2}\omega^2
 \left(c_2+c_3-\frac{g_A^2}{8m_N}\right)-{\mathbf{Q}}^2
 \left(1+\frac{4\hat{\rho}}{f^2}
 c_3-\frac{g_A^2 \hat{\rho}}{m_N f^2}\right)
 -\frac{g_A^2\hat{\rho}}{2f^2m_N}
 \frac{({\mathbf{Q}}^2)^2}{\omega^2}\nn \\
&&
 +\frac{g_A^2\bar{\rho}}{f^2 \omega}{\mathbf{Q}}^2-
 \frac{\bar{\rho} \omega}{f^2}=0~,
 \nn\\
 \mathbf{\pi^-:} 
&&\nn\\
&&
 \omega^2-M_{\pi^+}^2\left(1+c_1\frac{8\hat{\rho}}{f^2}\right)
 +\frac{4\hat{\rho}}{f^2}\omega^2
 \left(c_2+c_3-\frac{g_A^2}{8m_N}\right)
 -{\mathbf{Q}}^2\left(1+\frac{4\hat{\rho}}{f^2}
 c_3-\frac{g_A^2 \hat{\rho}}{m_N f^2}\right)
 -\frac{g_A^2\hat{\rho}}{2f^2m_N}
 \frac{({\mathbf{Q}}^2)^2}{\omega^2}\nn \\
&&
 -\frac{g_A^2\bar{\rho}}{f^2 \omega}{\mathbf{Q}}^2+
 \frac{\bar{\rho} \omega}{f^2}=0~,
\ea
with $\hat{M}_\pi^2=2 \hat{m} B_0$ the lowest order CHPT pion masses and
$M_{\pi^0}$, $M_{\pi^+}$ the vacuum pion masses at $\Opc$\cite{gl}. We will
take all of them to coincide with the corresponding physical masses since
differences will be $\Ops$.


\subsection{Pion dispersion and condensation}
\label{pion-dis-con}

The previous expressions largely simplify in the case 
of symmetric nuclear matter  $\barro=0$ in the chiral limit,
\be
 \omega^2={\mathbf{Q}}^2\left(1-\frac{4\hatr}{f^2}c_2\right).
\ee
The pion velocity in-medium for this case is given by:
\be
 \label{pv}
 \widetilde{v}=\frac{d \omega}{d |\mathbf{Q}|}=1-\frac{2\hatr}{f^2}c_2.
\ee 
Since it must be smaller than the velocity of light~\cite{leut,pt}, this
imposes the constraint $c_2\geq 0$ which is satisfied by the actual value of
this constant, $c_2=3.2\pm 0.25$ GeV$^{-1}$\cite{fettes}. Imposing
$\widetilde{v}\geq 0$, then $c_2\leq f^2/2\hatr$. Taking the previous value
for $c_2$ gives $\rho\leq 2\rho_0$. This clearly indicates that in-medium
CHPT, just by considering $\pi N$ interactions with the actual values of the
$c_i$ counterterms, can only be applied to densities $\lesssim \rho_0$, as
otherwise the corrections will be larger than $50\%$. For $\rho_0$ (nuclear
matter saturation density), $k_F \simeq 270\,$MeV$\sim 2 M_\pi$. Even for
vacuum $\pi N$ scattering, such a value for the running three-momentum is on
edge of applicability of the theory~\cite{fettes,fet2}.  However, one also has
to take into account that due to the large variation in the values of the
$c_i$ and combinations of them, important differences with respect to the
actual scale for a specific process are expected; see section \ref{breakdown}
for a discussion about the scales at which the chiral expansion breaks down.

We are now interested in finding solutions to eqs.\,(\ref{spec}) with
vanishing energy so that the ground state becomes unstable against the
spontaneous creation of pion modes. A close look to eqs.\,(\ref{spec})
indicates the presence of possible solutions in the antisymmetric matter case,
with ${\mathbf{Q}}^2/\omega={\rm constant}$ for 
$\omega,{\mathbf{Q}}\rightarrow 0$. To
be more specific, let us consider eq.\,(\ref{spec}) for negative pions
$\pi^-$.  Then for $\omega, {\mathbf Q}\rightarrow 0$ 
and ${\mathbf Q}^2/\omega=D$ one
has the equation: 
\be
 \label{pc1}
 \frac{g_A^2\hatr}{2 f^2 m_N}D^2+\frac{g_A^2 \barro}{f^2}D+M^2_{\pi^+}\left(
 1+\frac{8c_1 \hatr}{f^2}\right)=0
\ee
with the solution:
\be
 D=\frac{f^2 m_N}{g_A^2\hatr}\left[-\frac{g_A^2 \barro}{f^2}\pm \sqrt{
 \left(\frac{g_A^2 \barro}{f^2}\right)^2-\frac{2g_A^2\hatr}{f^2m_N}M_{\pi^+}^2
 \left(1+\frac{8c_1\hatr}{f^2}\right)   }\,\right].
\ee
In order to have an oscillatory mode and not a diffuse one, the following
inequality has to be fulfilled,
\be
 \label{pccon}
 \rho\geq \frac{4 f^2}{g_A^2 m_N}\left(\frac{\hatr}{\barro}\right)^2 
 M_{\pi^+}^2\left(1+\frac{8c_1 \hatr}{f^2}\right).
\ee
When the previous equality holds, $D=-m_N\,\barro/\hatr$. This implies that
for the $\pi^+$, where $D=m_N\,\barro/\hatr$, the energy is negative so that
it is energetically favorable to excite $\pi^+$ modes, i.e. we have a $\pi^+$
boson condensate. Because of this condensate, the U(1) flavor symmetry, which
is still present in eq.\,(\ref{fZ2}), is spontaneously broken, leading to one
Goldstone boson and two heavy modes because we started with three independent
pion modes. See ref.~\cite{son} for a similar discussion in the case of QCD at
finite isospin density $\mu_I$ in the idealized case of $\mu_B=0$.  Note also
that for the neutral pion there is no such solution.

For $|\barro|\ll\hatr$, it is justified to use eqs.\,(\ref{spec}) because the
expansion of the baryon propagators in eq.\,(\ref{pop}) still holds since then
$D\ll 2 m_N$.  However, when $|\barro| \sim \hatr$ the remarks discussed at
the end of sec.\,\ref{sec:counting} apply and a more careful treatment is
necessary.  Still, the prior situation is approximately realized in heavy
nuclei, where for $N/Z=1.5$ one has $\barro/\hatr = -1/5$. Thus we can apply
eq.\,(\ref{pccon}) with the result: 
\be
 \label{repchn}
 \rho\geq \frac{100 f^2}{g_A^2 m_N}M_{\pi^+}^2
 \left(1+\frac{4 c_1 \rho}{f^2}\right)\simeq
  8 \rho_0\left(1+\frac{4c_1 \rho}{f^2}\right),
\ee
so that:
\be
 \label{repchn2}
 \rho\left(1-\frac{32c_1 \rho_0}{f^2}\right)\geq 8\rho_0.
\ee
The previous inequality implies:
\be
 \label{pcaw}
 \rho\geq (1.6\pm 0.2)\rho_0~,
\ee
where we have used $c_1=-\left(0.81\pm 0.12\right)$\,GeV$^{-1}$ \cite{paul}. 
The resulting lower limit for the density in eq.\,(\ref{pcaw}) 
is larger than the total 
density in heavy nuclei, $\rho\simeq \rho_0$. 
Therefore, these oscillatory pion modes in heavy nuclei do not occur.

For neutron star matter, $\barro=-\hatr$, there is a reduction by a factor of
25 on the l.h.s. of eq.\,(\ref{repchn}) so that: \be \rho\geq \frac{4
  f^2}{g_A^2 m_N}M_{\pi^+}^2\left(1+\frac{4 c_1 \rho}{f^2}\right)\simeq 0.4
\rho_0\left(1+\frac{4c_1 \rho}{f^2}\right), \ee which amounts to a drastic
change. In fact, for this case, eq.\,(\ref{pop}) cannot be applied and one has
to take care of the modification of the counting by making use of
eq.\,(\ref{ffc}) case b) since here we have only pion lines.

In appendix \ref{sec:example} we discuss in detail the use of the non-standard
counting rules applied to the problem of the pion propagation in the medium.
These results will be used in the next section when determining the
appropriate in-medium scale in our framework.


\subsection{In-medium breakdown scales}
\label{breakdown}

In this section, we will deduce the in-medium scales at which the chiral
expansion breaks down.
To do so let us consider eq.\,(\ref{fZ2}) for the generating functional or
better its more general, non-perturbative expression given in
ref.~\cite{med1}. There, the in-medium contribution to the chiral Lagrangian
is given by: 
\be
 \label{fZ}
 \int d\vx\, d\vy \int^{k_F}\!\!\frac{d\vp}{(2\pi)^3 2E(p)} 
 e^{-i\vp (\vx-\vy)}\hbox{Tr}
 \left\{(\barr{p}+m_N)\log {\cal F} \right\}~,
\ee
with $\log {\cal F}$ being a pure interaction operator whose perturbative
expansion leads to eq.\,(\ref{fZ2}) (for definitions, see ref.~\cite{med1}).
In the limit $k_F\rightarrow 0$, the leading non-relativistic contribution, if
not vanishing, comes from $\barr{p}+m_N \rightarrow 2\,m_N$ and the above
expression can be written as 
\be
 \label{scala}
 \int d\vx\, d\vy \hbox{Tr}\left\{2m_N\,\log{\cal F} \right\} \int^{k_F}\!\! 
 \frac{d\vp}{(2\pi)^3 2m_N}
 =\int d\vx \frac{k_F^3}{6\pi^2}\left(\int d\vy \hbox{Tr}
 \left\{\log {\cal F}\right\}\right),
\ee
involving the numerical factor $1 / 6\pi^2$ and establishing that the
in-medium corrections must start at order $k_F^3\propto \rho$. Taking into
account the perturbative expansion of $\log {\cal F}$, given in
eq.\,(\ref{fZ2}) \cite{med1} and noting that the local operators $A$ have
dimension of mass, see e.g. eqs.~(\ref{a1}) and (\ref{pin2}), one concludes
that any term arising from eq.~(\ref{scala}) with $m$ pion fields and $n$
momenta, $n\geq 1$, counts as: 
\be
 \label{dan}
 \frac{k_F^3}{6\pi^2}\frac{\phi^m}{f^m}\frac{Q^n}{\Lambda^{n-1}},
\ee
with $Q^n$ involving both derivatives and quark mass insertions and $\Lambda$
refers to a typical hadronic mass $\sim 1$ GeV. On the other hand from
$\La_{\pi\pi}$ one typically has, $n\geq 2$: 
\be
 \label{dan2}
 \frac{\phi^m}{f^{m-2}}\frac{Q^n}{\Lambda^{n-2}}.
\ee
To compare  eqs.\,(\ref{dan}) and (\ref{dan2}) it is convenient to rewrite 
the former as:
\be
 \label{dan3}
 \frac{k_F^2}{6\pi^2 f^2}\frac{\phi^m}{f^{m-2}}\frac{Q^n k_F}{\Lambda^{n-1}}.
\ee
Since in our counting $k_F\sim Q$ we should consider $n\rightarrow n-1$,
$n\geq 2$, in eq.\,(\ref{dan3}) so that the local in-medium contributions are
suppressed by $k_F^2/(6\pi f^2)$ as compared to the vacuum ones. It is also
worthwhile to stress that while the leading in-medium contribution scales as
$k_F^3\sim \rho$ when $k_F \rightarrow 0$, eq.\,(\ref{scala}), the suppression
with respect to a large scale surviving in the chiral limit, much larger than
pion mass or momenta, involves the second and not the third order in $k_F$,
i.e. the suppression is non-analytic in the nuclear density.  Note that this
scale, $\sqrt{6} f \simeq 700\,{\rm MeV}$, can also be inferred from the
S-wave Weinberg-Tomozawa contribution (the last term in eqs.(\ref{spec}) for
the charged pions).  Under $P$-wave interaction, see the last but one term in
eqs.\,(\ref{spec}) for the charged pions, the scale is reduced to $\sqrt{6}
f/g_A \simeq 570\,{\rm MeV}$, a little bit larger than twice the Fermi
momentum for nuclear matter saturation density.  It is roughly a factor of two
smaller than the commonly used value $4\pi f_\pi$ typically quoted in vacuum
CHPT for the pion sector. This is similar to the flavor dependence $\sim
1/{N_f}$ pointed out in the context of extended technicolor approaches to
electroweak symmetry breaking \cite{soldate}.

\begin{figure}[thb]
\centerline{\epsfig{file=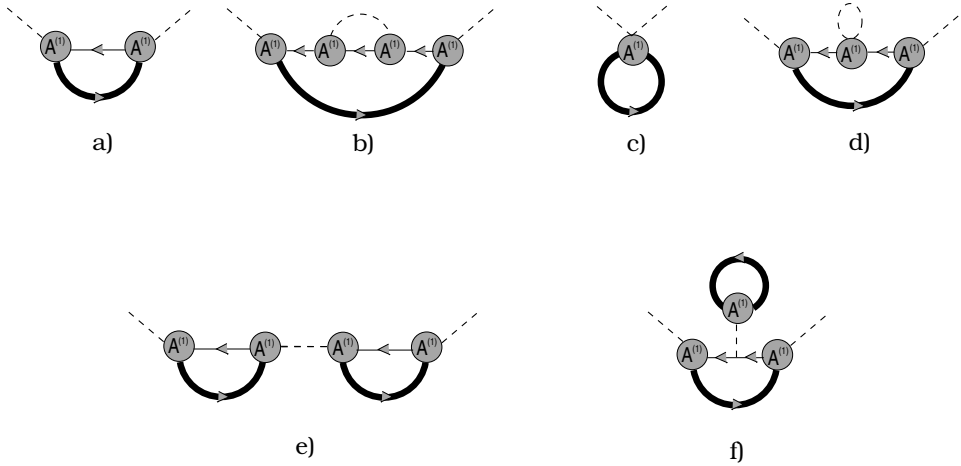,width=5.in}}
\vspace{0.3cm}
\caption[pilf]{\protect \small 
  Diagrams that contribute to the in-medium propagation of pions up to $\Opc$
  when $Q^0\sim k_F^2/2m_N$, eq.\,(\ref{ffc}) case b). Diagram a) is $\Opt$
  and the rest are $\Opc$. Other diagrams obtained by exchanging free baryon
  propagators by Fermi-sea insertions or by exchanging lines and or the
  position of loops are not shown.
\label{fig:pp2}}
\end{figure}

The main problem when $Q_j^0\sim Q^2/2m_N$, i.e.\ the non-standard scenario
for the counting rules, comes from the further reduction in the scale
$\Lambda$ of the nuclear perturbative expansion pertinent to such cases.  In
fact, instead of having the factor $2 m_N Q^0$ as in eq.\,(\ref{pop}), one
needs to consider the full expression since $Q^0$ counts now as a quantity of
order $k_F^2/2m_N$ in the denominator, $k_F\sim \Op$. As a consequence, if
something counted before as $k_F^2/\Lambda^2$, it now scales as $2 m_N
k_F/\Lambda^2$. This implies that $\Lambda \rightarrow \Lambda^2/2m_N \simeq
170$ MeV instead of $\Lambda\simeq 570$ MeV, that is, a reduction by roughly a
factor of three. For instance, considering eq.\,(\ref{specg}) for the
propagation of pions in the case of $\omega=0$ 
we have from fig.\,\ref{fig:pp2}a,
which in fact corresponds to figs.\ref{fig:pp}b,c: 
\be
 \label{newfu}
 \frac{2 m_N g_A^2 {\mathbf Q}^2}{f_\pi^2}\left[\frac{k_F}{8\pi^2}+
 \frac{4k_F^2-{\mathbf Q}^2}{32 \pi^2 |\mathbf{Q}|}\log 
 \frac{2k_F+|\mathbf Q|}{|2k_F-|\mathbf{Q}||} \right]~.
\ee
Taking now $|{\bf{Q}}|\ll k_F$, the previous expression reduces to:
\be
 \label{ns}
 {\bf Q}^2\frac{m_N g_A^2 k_F}{2\pi^2 f_\pi^2}~,
\ee
with the scale $\Lambda'=4\pi^2 f_\pi^2/2m_N g_A^2\sim 120~\hbox{MeV}$, of the
size we have estimated just above. As a result, while for the standard
counting scenario one has a chiral expansion to be applied when $k_F\lesssim 2
M_\pi$, corresponding to nuclear saturation density, when $Q_j^0\sim Q^2/2m_N$
the perturbative expansion can only be applied up to densities around $\sim
\left(170~\hbox{MeV}/k_F^0\right)^3\simeq \rho_0/4$, where $k_F^0\simeq 268$
MeV is the Fermi momentum at nuclear matter saturation density. This
modification of the scale is simply lost in the mean-field approaches to
nuclear matter.  However, as we have seen in sec.\,\ref{pion-dis-con} and
appendix\,\ref{sec:example}, this scale is crucial for the study of pion
condensation.

For the case of the $K^-$ condensation in nuclear matter studied in
\cite{kaplan} under the use of the mean-field approach, the absence of baryons
with strangeness $+1$ implies that an analogous diagram to that of
fig.\,\ref{fig:pp2}a is not involved.

The previous problem is similar to that of studying nucleon-nucleon
interactions in the vacuum where one also has this situation~\cite{wein} due
to the appearance of the large $2m_N \sim {\cal O}(1)$ factor instead of an
$\Op$ one. The solution advocated in this case is to perform a resummation of
diagrams by solving the Lippmann-Schwinger equation~\cite{wein} and resumming
those diagrams with two nucleons intermediate states. It is clear that one
should proceed in a similar way here by resumming an infinite subset of
diagrams, including as well nucleon contact interactions, as, on the other
hand, is known for a long time in many-body problems~\cite{fett,mat}. More
specifically, including extra $\rho-$vertices repeatedly iterated as in
fig.\,\ref{fig:pp2}e, with any of them enhanced by the appearance of the large
factor $2m_N/Q$, amounts to the RPA approximation for the in-medium pion
propagator.


\subsection{In-medium pion mass}
\label{subsec:pim} 

There has been recent interest in calculating the pion masses in the medium
using heavy baryon chiral perturbation theory (HBCHPT) \cite{ko,kw1}.  Both
works include one pion loop graphs, $\Ops$, and while ref.~\cite{ko} is
restricted to the nuclear symmetric case, ref.~\cite{kw1} also considered
non-symmetric nuclear matter although, as shown below, not all the effects of
order $\Ops$ are included. However, the subject has already a long history and
similar diagrams than those calculated in refs.~\cite{ko,kw1} were already
considered in earlier (chiral) models \cite{eu}, and the pole position of the
pion propagator in nuclear matter has been discussed in the context of pion or
kaon condensation, see e.g. \cite{pole}. The calculation of the in-medium pion
mass implies to impose in eqs.\,(\ref{spec}) the condition $\bf{Q}=0$ and then
to solve for $\omega=\widetilde{M}_\pi$, where the tilde denotes in-medium pion
mass (afterwards, when discussing a QCD Ward identity and pion condensation,
the full spectral relations will turn out to be essential).
{}From eqs.\,(\ref{spec}), the 
equations to determine $\widetilde{M}_\pi$ are
\ba
 \label{pim}
 \pi^0: \quad
&&
 \widetilde{M}_{\pi^0}^2-M_{\pi^0}^2\left(1+c_1\frac{8\hat{\rho}}{f^2}\right)
 +\frac{4\hatr}{f^2}\widetilde{M}^2_{\pi^0}
 \left(c_2+c_3-\frac{g_A^2}{8m_N}\right)
 -M_\pi^2\frac{4\bar{\rho}}{f^2}\frac{\bar{m}}{\hat{m}}c_5=0~,\nn\\
 \pi^+: \quad
&&
 \widetilde{M}_{\pi^+}^2-M_{\pi^+}^2\left(1+c_1\frac{8 \hatr}{f^2}\right)+
 \frac{4\hatr}{f^2}\widetilde{M}_{\pi^+}^2
 \left(c_2+c_3-\frac{g_A^2}{8m_N}\right)
 -\frac{\bar{\rho}\widetilde{M}_{\pi^+}}{f^2}=0~,
\nn\\
 \pi^-: \quad
&&
 \widetilde{M}_{\pi^-}^2-M_{\pi^-}^2\left(1+c_1\frac{8\hat{\rho}}{f^2}\right)+
 \frac{4\hatr}{f^2}\widetilde{M}_{\pi^-}^2
 \left(c_2+c_3-\frac{g_A^2}{8m_N}\right)
 +\frac{\barro \widetilde{M}_{\pi^-}}{f^2}=0~.
\ea
Up to the order we are considering here it is correct to substitute
$\widetilde{M}_\pi \rightarrow M_\pi$ in all the expressions above except when
$\widetilde{M}_\pi^2$ appears alone in the beginning of these equations. We
will specialize our study to the case of $^{207}$Pb, a heavy nucleus, where
deeply bound $\pi^-$ states have been detected \cite{g,i} with a shift in the
effective in-medium $\pi^-$ mass $\Delta M_{\pi^-}=23-27$ MeV \cite{i} under
the use of pion-nucleus optical potentials. To facilitate the comparison with
ref.~\cite{kw1} we take the same values of $k_p=246.7$ MeV and $k_n=282.4$ MeV
corresponding to almost nuclear matter density saturation and $N/Z=1.5$. These
are typical values for heavy nuclei.

As values for the next-to-leading order pion-nucleon counterterms we take from
ref.~\cite{paul} $c_1=-0.81\pm 0.12$ GeV$^{-1}$, $c_3=-4.70\pm 1.16$
GeV$^{-1}$, whereas $c_2$ is essentially undetermined by this analysis. The
value of $c_3$ has been also determined in ref.~\cite{rent} in good agreement
with the previous value. Finally from ref.~\cite{fettes,fet2} we take
$c_2=3.2\pm 0.25$ GeV$^{-1}$. A look at eqs.\,(\ref{pim}) reveals that the
previous formulae only depend on the combination $c_2+c_3$ which can be pinned
down more precisely from the experimental value of $T^+(M_\pi)=-0.045\pm
0.088\,$fm~\cite{sch}. Following ref.~\cite{bkm1} one has: 
\be
 \label{pindown}
 T^+(M_\pi)
 =\frac{M_\pi^2}{f^2}\left(-4c_1+2c_2-\frac{g_A^2}{4m_N}+2c_3\right)+\Opt.
\ee
Considering the previous value for $c_1$ one finally has $c_2+c_3=-1.46 \pm
0.26$ GeV$^{-1}$ with a central value in agreement with the values for $c_2$
and $c_3$ quoted above. In fact, taking the value of $c_2$ from \cite{fettes}
it follows that $c_3=-4.66\pm0.36$ GeV$^{-1}$, which is the value we will
finally take. It is more precise than that of ref.~\cite{paul} but with the
same central value. {}From the previous equation we can also determine the
combination $-2c_1+c_2+c_3=0.16\pm 0.10$ GeV$^{-1}$.  This result is
consistent within errors with the one obtained from analyzing pion--deuteron
scattering in chiral perturbation theory, $-2c_1+c_2+c_3=-0.04\pm 0.02$
GeV$^{-1}$~\cite{bblm}. The difference in the central value reflects an $\Opt$
contribution to $T^+(M_\pi)$\cite{bkm1}. Nevertheless, when this combination
of counterterms appears we will directly take the empirical value of
$T^+(M_\pi)$.

\renewcommand{\arraystretch}{1.2}
\begin{table}[h]
\begin{center}
\begin{tabular}{|c|rrr|}
\hline
& $Exact~(\Delta M)$ & $Exact~(M)$ & $Perturbative$ \\
\hline
$\Delta M_{\pi^-}$ & 18 $\pm$ 5 MeV   & 18 $\pm$ 16 MeV    
  & 8.2 $\pm$ 2 MeV \\
$\Delta M_{\pi^+}$ & $-12$ $\pm$ 4 MeV & $-12$ $\pm$ 13 MeV
 & $-6.5$ $\pm$ 2 MeV\\ 
$\Delta M_{\pi^0}$ & 2 $\pm$ 4 MeV   & 2 $\pm$ 7.0 MeV     
 & 1.1 $\pm$ 2 MeV\\
\hline
\end{tabular}
\end{center}
\centerline{\parbox{13cm}{
\caption{
 Shift between the in-medium pion effective masses and the vacuum ones. 
 The label $Exact~(\Delta M)$  refers 
 to solve eqs.\,(\ref{pim}) exactly in terms of 
 $\delta \omega^2\equiv \widetilde{M}_\pi^2-M_\pi^2$ and analogously 
 for $Exact~(M)$ but directly for $\widetilde{M}_\pi$. 
 Finally, {\it Perturbative} 
 means to solve the eqs.\,(\ref{pim}) by 
 replacing $\widetilde{M}_\pi$ by $M_\pi$ in all 
 the terms with density dependence.
\label{tab:pmm}}}}
\end{table}

\noindent
In the third column of table~\ref{tab:pmm} we show our numerical results for
$^{207}$Pb in the perturbative case when substituting in eq.\,(\ref{pim})
$\widetilde{M}_\pi\rightarrow M_\pi$ in all the terms with density dependence.
As we can see the resulting shift in the mass of the $\pi^-$ is much smaller
than the experimental one, $\Delta M_{\pi^-} = 23-27$ MeV.  However, looking
closely at eq.\,(\ref{pim}) one realizes the presence of large in-medium
corrections, of the order of 50$\%$. These originate from the terms within the
round brackets, namely $1+4\hatr(c_2+c_3-g_A^2/8m_N)/f^2\simeq 1+8\hatr
c_1/f^2 \simeq 0.5$ for nuclear matter saturation density. Thus doubts arise
with respect to the accuracy of eqs.\,(\ref{pim}) indicating that higher order
corrections will give non-negligible contributions. However, this also
indicates that if one gives credit to eqs.\,(\ref{pim}) for such densities,
one has to solve them exactly. Otherwise one is ignoring large corrections to
the solutions of eqs.\,(\ref{pim}) when solving them in a pure perturbative
way.  Note that if $\delta=-0.5$ such that $1+\delta=0.5$, then the deviation
from one for ratios like $1/(1+\delta)$, appearing when solving
eqs.\,(\ref{pim}), is a $100\%$ effect. On the other hand, solving
eqs.\,(\ref{pim}) in terms of $\widetilde{M}_\pi$, as they are written, one
ends with the results collected in the second column of table~\ref{tab:pmm}.
When comparing columns two and three two important points are easily noted: 1)
The sizeable difference, more than a factor of two, in the central value of
the shift of the $\pi^-$ mass when comparing the full and perturbative
solutions and 2) the errors for the second column are so large that they
prevent one from obtaining any definitive conclusion about the mass shift.
This situation can be improved if one rewrites eqs.\,(\ref{pim}) in terms of
$\delta \omega^2$ defined by the change of variables
$\widetilde{M}^2_\pi=M_\pi^2+\delta \omega^2$ 
and then solving for $\delta \omega^2$ in
the $\pi^-$, $\pi^+$ and $\pi^0$ cases. In this way one is sensitive to
$T^+(M_\pi)\rho$ and to $c_2+c_3$. Furthermore, the square of the vacuum pion
mass is removed in the equations, which sizably decreases the sensitivity
with respect to $c_2+c_3$. Proceeding in this way one obtains the numbers in
the first column of table~\ref{tab:pmm} that constitute our final results.
Within errors, these are compatible with the experimental interval of $\Delta
M_{\pi^-}=23-27\,$MeV.

Higher order corrections to this result have to be evaluated but not using
$\widetilde{M}_\pi=M_\pi$ as it was done in ref.~\cite{kw1}.  Indeed, we also
indicate here that the corrections to $M_\pi^2$ in an asymmetric nuclear
medium, because of the Weinberg term (the last terms in eqs.\,(\ref{pim}) for
the charged pions), are $\Opc$. Hence if one is interested in an $\Ops$
calculation as in ref.~\cite{kw1}, one should take the value at $\Opc$ for
$\widetilde{M}_\pi$ and use this one to determine once again the Weinberg
term. If this is not done, one loses $\Ops$ contributions which in fact were
not taken into account in ref.~\cite{kw1} because there the in-medium
self-energy was calculated with $Q=(M_\pi,\mathbf{0})$. However, such kind of
corrections are numerically small as they contribute just $\sim 0.5$ MeV to
$\Delta M_{\pi^-}$.

To conclude, our results in the first column in table~\ref{tab:pmm}
tentatively indicate that in-medium CHPT accounts for most of the required
shift in the measured $\pi^-$ mass at finite density from recent experiments
on deeply bound pionic atoms \cite{i}. The main contribution to this shift,
around 16 MeV, results from the division of the Weinberg term by the factor
$1+4\hatr(c_2+c_3-g_A^2/8m_N)/f^2$. The latter factor corresponds to the
wave-function renormalization of the in-medium pions in symmetric nuclear
matter at threshold, c.f. eq.\,(\ref{z}), which for nuclear saturation density
is $\sim 0.5$. This last point, related to the propagation of pions in the
medium, was not accounted for in ref.~\cite{kw1}. Notice as well that the
first term in brackets in eqs.\,(\ref{pim}) is also around 0.5, such that,
when divided by the wave-function renormalization constant, it is slightly
greater than one and it further increases $M_{\pi^-}$ by the small amount of 2
MeV, the same value appearing in the $M_\pi^0$ case, giving finally the 18 MeV
of increase for the $\pi^-$ case reported in table~\ref{tab:pmm}.


\subsection{Axial-vector and pseudoscalar pion couplings}
\label{subsec:fps}

The in-medium axial-vector current $A^i_\mu(x)=\bar{q}(x)\gamma_\mu \gamma_5
(\tau^i /2) q(x)$ is given at the tree-level by a calculation of the
functional derivative of the classical action $\int dy 
\widetilde{{\cal  L}}[a,v,s,p]$ 
with respect to $a_\mu^i(x)$ where $a(x)=\sum_i
a_\mu^i(x)\tau^i/2$ and where, in the end, the limit of vanishing external
sources is taken. In a diagrammatic language and for the in-medium
contributions, this amounts to consider analogous diagrams to those of
fig.\,\ref{fig:pp} but with one external pion line substituted by an
axial-vector source. In addition, one needs to consider the pion wave function
renormalization. This deserves special attention in our case because the
non-relativistic and non-local character of the in-medium chiral Lagrangian
deduced in ref.\cite{med1}, as also discussed in
appendix~\ref{subsec:genfunc}, eq.\,(\ref{fZ2}). Thus standard prescriptions
to obtain the wave function renormalization are not appropriate here and a
detailed calculation of the latter is given in appendix \ref{subsec:wfr}.

We also introduce the axial-vector current $J_{A\,\mu}^{12}=A^{1}_\mu-i
A^{2}_\mu\equiv \bar{d}\gamma_\mu\gamma_5 u$ and its hermitian conjugate
$(J_{A\,\mu}^{12})^\dagger$.  Furthermore, because of the breaking of
covariance due to the presence of the nuclear medium, it is convenient to
separate between temporal and spatial couplings of the pions to the
axial-vector currents.  In this way we have: 
\ba
 \label{caf}
 \langle \Omega| A_0^k|\pi^0\rangle
 &=&
 i\delta^{k3} f_\pi Q_0 \left\{
 1+\frac{2\hatr}{f^2}
 \left( c_2+c_3-\frac{g_A^2}{8m_N}\frac{M_\pi^2}{Q_0^2}\right)
 \right\} ~, \nn\\
 \langle \Omega| A_i^k|\pi^0\rangle&=&i\delta^{k3} f_\pi Q_i \left\{
 1-\frac{2\hatr}{f^2}\left(c_2-c_3+\frac{g_A^2}
 {8m_N}\frac{M_\pi^2}{Q_0^2}\right)
 \right\}~, \nn\\
 \langle \Omega|J_{A\,0}^{12}|\pi^+\rangle &=& i\sqrt{2} f_\pi Q_0\left\{1+
 \frac{2\hatr}{f^2}
 \left(c_2+c_3-\frac{g_A^2}{8m_N}\frac{M_\pi^2}{Q_0^2}\right)-
 \frac{3\barro}{4f^2 Q_0}-\frac{g_A^2\barro \vQ^2}{4f^2 Q_0^3}\right\}~, \nn\\
 \langle \Omega|J_{A\,i}^{12}|\pi^+\rangle&=&i \sqrt{2}f_\pi Q_i\left\{
 1-\frac{2\hatr}{f^2}\left(c_2-c_3+\frac{g_A^2}
 {8m_N}\frac{M_\pi^2}{Q_0^2}\right)+
 \frac{\barro}{4f^2 Q_0}-\frac{g_A^2\barro}
 {4f^2 Q_0}\left(4+\frac{\vQ^2}{Q_0^2}\right)
 \right\}~,\nn\\
 \langle \Omega|(J_{A\,0}^{12})^\dagger|\pi^-\rangle
 &=&
 i\sqrt{2}f_\pi Q_0\left\{1+
 \frac{2\hatr}{f^2}
 \left(c_2+c_3-\frac{g_A^2}{8m_N}\frac{M_\pi^2}{Q_0^2}\right)+
 \frac{3\barro}{4f^2 Q_0}+\frac{g_A^2\barro \vQ^2}{4f^2 Q_0^3}\right\}~, \nn\\
 \langle \Omega|(J_{A\,i}^{12})^\dagger|\pi^-\rangle
 &=&
 i\sqrt{2} f_\pi Q_i\left\{
 1-\frac{2\hatr}{f^2}\left(c_2-c_3+\frac{g_A^2}
 {8m_N}\frac{M_\pi^2}{Q_0^2}\right)-
 \frac{\barro}{4f^2 Q_0}+\frac{g_A^2\barro}
 {4f^2 Q_0}\left(4+\frac{\vQ^2}{Q_0^2}\right)\right\},
\ea
where $f_\pi=92.4$ MeV is the vacuum weak pion decay constant. Notice that the
terms in curly brackets for the temporal components correspond to the square
root of the wave function renormalization, ${\cal Z}(Q)^{1/2}$,
eq.\,(\ref{wfr2}), only in the case of symmetric nuclear matter.  Furthermore,
even in this case the inverse pion propagator, obtained from the l.h.s of
eqs.\,(\ref{spec}), is not equal to 
$Z_\alpha(\vQ^2)(\omega^2-\omega^2_\alpha(\vQ^2))$
(where $\omega_\alpha({\bf Q}^2)$ corresponds to the positive pion energies of
isospin state $\alpha$ as given by eqs.\,(\ref{spec})) in an unambiguous way,
since $Z_\alpha$ has energy dependence determined only on-shell.
Nevertheless, this last effect is suppressed in the non-relativistic limit.
Indeed, $c_2+c_3$ is about one order of magnitude larger than $g_A^2/8m_N$.

Let us now introduce the couplings $f_t^0$ (temporal) and $f_s^0$ (spatial) as
the value of the matrix elements $\langle \Omega |A^3_0|\pi^0\rangle/i Q_0 $
and $\langle \Omega| A^3_i|\pi^0\rangle/iQ_i$ at threshold, i.e. for vanishing
three--momentum.  In the same way we also define the couplings $f_t^{+}$ and
$f_s^+$ as the matrix elements $ \langle
\Omega|J_{A\,0}^{12}|\pi^+\rangle/i\sqrt{2}Q_0$ and $\langle
\Omega|J_{A\,i}^{12}|\pi^+\rangle/i\sqrt{2}Q_i$ at threshold and analogously
for the $\pi^-$ with the couplings $f_t^-$ and $f_s^-$, that can be obtained
from the former just by exchanging $\barro \rightarrow -\barro$.

In the isospin limit, $\bar{m}=\barro=0$, the  weak couplings for all the 
pions are equal and given by:
\ba
\label{fs}
f_t&=&f_\pi\left\{1+
\frac{2\hatr}{f^2}\left(c_2+c_3-\frac{g_A^2}{8m_N}\right)\right\}~,\nn\\
f_s&=&f_\pi\left\{
1-\frac{2\hatr}{f^2}\left(c_2-c_3+\frac{g_A^2}{8m_N}\right)\right\}~,
\ea
see also ref.\cite{aw}. {}From these equations we see that the spatial
coupling, $f_s$, is suppressed with respect to the temporal one by a factor:
\be
\label{fsup}
\widetilde{v}^2=1-\frac{4\hatr c_2}{f^2},
\ee
since $c_2$ is positive as already discussed. Remember that $\widetilde{v}$,
given in eq.\,(\ref{pv}), corresponds to in-medium pion velocity in symmetric
nuclear matter in the chiral limit.  Taking into account the values for the
low-energy constants given in sec.~\ref{subsec:pim}, we can write for $f_t$
and $f_s$: 
\ba
 \label{ft}
 f_t&=&f_\pi\left\{1-\frac{\rho}{\rho_0}(0.26\pm 0.04)\right\}~,\nn\\
 f_s&=&f_\pi\left\{1-\frac{\rho}{\rho_0}(1.23\pm 0.07)\right\}~,
\ea
with $\rho_0$ the nuclear saturation density. We see that the reduction with
density is sizeable for $f_t$, being $1/4\,f_\pi$ at $\rho_0$, and dramatic
for $f_s$, which vanishes already for $\rho\simeq 0.8\rho_0$. We remark that
in the vacuum, the study of the left-right current correlator leads to the
conclusion that spontaneous chiral symmetry breaking is unambiguously signaled
by the non--vanishing of the pion decay constant $f_\pi$. By considering the
coupling of the pion to the axial-vector {\it charges} $Q_A^k=\int d
\vx\,A_0^k(\mathbf{x})$, the generators of the axial chiral transformations at
the quantum level, one can see that it is the temporal component $f_t$ that
plays the role of $f_\pi$ in the medium.  Indeed, a straightforward
calculation taken into account eqs.\,(\ref{caf}) shows: 
\be
\label{chargp}
\langle\pi^a(\vp)|Q_A^k|\Omega\rangle
 =\int d\vx\,e^{-i\vp\vx}\langle\pi^a|A_0^k(0)|
\Omega\rangle=(2\pi)^3\delta(\vp)iQ_0\, f_t~.
\ee
Mathematically, one can obtain a more meaningful result than
eq.\,(\ref{chargp}), that involves the product $\delta(\vp)iQ_0$ which in the
chiral limit needs some care, by considering the wave packet: 
 \be \int
 \frac{d\vp}{(2\pi)^3 2|\vp|} f(\vp^2)|\pi^a(\vp)\rangle~, 
\ee 
with $f(0)={\rm constant}$ as discussed in ref.~\cite{gws}. Thus, as long as
$f_t\neq 0$ the vacuum is not invariant under axial chiral transformations and
one has for symmetric nuclear matter the same chiral symmetry breaking pattern
as in vacuum, namely, $SU(2)_L\times SU(2)_R \rightarrow SU(2)_V$. Of course,
the fact that $f_s \neq f_t$ in matter (and some consequences thereof) was
already discussed in the literature, see e.g. \cite{leut,kr,aw,aw2,pt}.  In
addition, it should also be interesting to perform the in-medium analysis of
the left-right current correlator as in the vacuum.

One can give a more microscopic picture for the large quenching of the spatial
components of the axial-vector current by considering the resonance saturation
of the $\Opd$ CHPT counterterms, the $c$'s, making use of ref.~\cite{bkm1}. In
this way we also connect with earlier many-body approaches~\cite{rho,tow},
which stress the importance of the isobar $\Delta(1232)$ in the related
quenching of Gamow-Teller transitions in finite nuclei. In fact we agree with
this observation since the $\Delta$ contributes about $80\%$ to the very large
number $c_2-c_3=7.9$ GeV$^{-1}$ which controls the in-medium quenching of
$f_s$, eq.\,(\ref{ft}), while the other main contribution, around a $15\%$, is
due to the crossed exchange of a heavy scalar resonance
\cite{bkm1,pin}\footnote{Although in ref.~\cite{bkm1} this contribution was
  mimicked by the exchange of a light $\sigma$, in ref.~\cite{pin} it was
  pointed out that when chiral symmetric Lagrangians are used and unitarized
  in harmony with CHPT to describe $\pi N$ scattering, then the resulting
  crossed scalar contribution comes out to be heavy.}. On the contrary, the
contribution of the $\Delta$ to $f_t$ vanishes because then the sum $c_2+c_3$
appears. In this case the main contributions arise equally from the
$N^*(1440)$ and the crossed scalar exchange and amount to a much smaller
quantity than the difference $c_2-c_3$ as shown in eq.\,(\ref{ft}).

Now, let us take into account eqs.\,(\ref{fs}) and the resulting expression,
from eqs.\,(\ref{pim}), of the in-medium pion mass in the $\rho-$isospin
limit, just defined above: 
\be
 \label{pim2}
 \widetilde{M}_\pi^2=M_\pi^2\left\{1+\frac{4\hatr}{f^2}\left(2c_1-c_2-c_3+
 \frac{g_A^2}{8m_N}\right)\right\},
\ee
which, together with eq.\,(\ref{qc}) for the $\rho-$quark condensate $\langle
\Omega|\bar{u}u+\bar{d}d|\Omega \rangle$, establishes that in-medium
corrections up to the next-to-leading order for the case of symmetric nuclear
matter do not destroy the validity of the Gell-Mann-Oakes-Renner relation
(GMOR), which is only exact at lowest order: 
\be
 \label{gmor}
 f_t^2 \widetilde{M}_\pi^2
 =-\hat{m}\langle\Omega|\bar{u}u+\bar{d}d|\Omega\rangle+\delta_{\rm vac}~,
\ee
where here $\delta_{\rm vac}$ is a vacuum CHPT correction at $\Opc$.  The
stability of the GMOR relation under in-medium corrections as well as the fact
that it is the {\em temporal} coupling $f_t$, and not $f_s$, that is the one
involved in the GMOR relation, has been previously reported in
refs.~\cite{aw,aw2}, within the mean field approximation, and in
ref.~\cite{yon}, in the framework of QCD sum rules.

We now consider the general case with $\barro\neq 0$ and turn our attention to 
the pseudoscalar isovector quark current
\be
 P^i=\bar{q}i\gamma_5 \tau^i q~,
\ee
which can be obtained following similar steps as in the case of the
axial-vector currents by differentiating the generating functional with
respect to $p^i(x)$ with $p(x)=\sum_i p^i \tau^i$, taking finally the limit of
vanishing external sources. We are interested in evaluating $\langle \Omega
|P^i|\pi^k\rangle$. Since the pseudoscalar source counts as $\Opd$ it first
appears in $A^{(2)}$ and one has to evaluate an analogous diagram to that of
fig.\,\ref{fig:tad} with the wavy line indicating now a pseudoscalar source.
It is then straightforward to obtain: 
\ba
 \label{pc}
\langle \Omega|P^i|\pi^3\rangle&=&G_\pi \delta^{i3}\left\{1+\frac{2\hatr}{f^2}
\left(4c_1-c_2-c_3+\frac{g_A^2}{8m_N}\frac{M_\pi^2}{Q_0^2}\right)\right\}~,
 \nn\\
\langle \Omega|P^{12}|\pi^+\rangle&=&\sqrt{2}G_\pi \left\{1+\frac{2\hatr}{f^2}
\left(4c_1-c_2-c_3+\frac{g_A^2}{8m_N}\frac{M_\pi^2}{Q_0^2}\right)+
\frac{\barro}{4f^2 Q_0}-\frac{g_A^2 \barro \vQ^2}{4f^2 Q_0^3}\right\}~,\nn\\
\langle \Omega|(P^{12})^\dagger|\pi^-\rangle&=&\sqrt{2}G_\pi 
 \left\{1+\frac{2\hatr}{f^2}
\left(4c_1-c_2-c_3+\frac{g_A^2}{8m_N}\frac{M_\pi^2}{Q_0^2}\right)
-\frac{\barro}{4f^2 Q_0}+\frac{g_A^2\barro \vQ^2}{4f^2Q_0^3}\right\}~,
\ea
with $G_\pi$ the vacuum pseudoscalar pion coupling calculated at $\Opc$
\cite{gl}.

This leads naturally to the study of the Ward identity relating the divergence
of the axial-vector quark current with the pseudoscalar ones,
\be
\label{wi}
\partial^\mu \left(\bar{q}\gamma_\mu \gamma_5 \frac{\tau^i}{2}q\right)=
\hat{m}\bar{q}i\gamma_5 \tau^i q +\, \delta^{i3} \bar{m}\bar{q}i \gamma_5 q~.
\ee 
Setting $\bar{m}=0$ and evaluating this expression at threshold one obtains a
relation between $f_t^{0,+,-}$, $\widetilde{M}_{\pi^{0,+,-}}^2$ and
$\widetilde{G}^0_\pi$, $\widetilde{G}^+_\pi$ and $\widetilde{G}^-_\pi$. The
latter are defined in terms of the matrix elements eq.\,(\ref{pc}) at
threshold, when the factor $\sqrt{2}$ is removed in the charged matrix
elements.  Considering first the $\pi^0$ case and from eqs.\,(\ref{caf}) and
(\ref{pc}), it follows: 
\ba
 \label{derv}
 \partial^\mu \langle \Omega|A^3_\mu(0)|\pi^0(Q)\rangle&=&f_t^0 
 \widetilde{M}_{\pi^0}^2=f_\pi \widetilde{M}_{\pi^0}^2
 \left\{1+\frac{2\hatr}{f^2}\left(c_3+c_2-\frac{g_A^2}{8m_N}
 \right)\right\}\nn\\
 &=&f_\pi M_\pi^2\left\{1+\frac{2\hatr}{f^2}\left(4c_1-c_2-c_3+
 \frac{g_A^2}{8m_N}\right)\right\}=\hat{m}\widetilde{G}_\pi~,
\ea
since of course in the vacuum $f_\pi M_\pi^2=\hat{m}G_\pi$ holds \cite{gl}.
Then
\be
 \label{relpa0}
 f_t^0 \widetilde{M}^2_{\pi^0}=\hat{m}\widetilde{G}_\pi~,
\ee
see also ref.\cite{aw}. An analogous relation for the charged pions can be
also obtained by considering the divergence of the axial current $J_A^{12}$
together with the pseudoscalar current $P^{12}$. One can readily check that
\ba
 \label{relpch}
 f_t^+ \widetilde{M}^2_{\pi^+}&=&\hat{m}\widetilde{G}_\pi~, \nn \\
 f_t^- \widetilde{M}^2_{\pi^-}&=&\hat{m}\widetilde{G}_\pi~.
\ea
It is worthwhile to stress that in eqs.\,(\ref{relpa0}) and (\ref{relpch}) we
have not imposed that $\barro=0$ but just $\bar{m}=0$ as is necessary in order
that the Ward identity of eq.\,(\ref{wi}) holds in terms of only the isovector
pseudoscalar currents.

We now consider the general case with $\bar{m}\neq 0$. To study
eq.\,(\ref{wi}), it is necessary to consider as well the isoscalar
pseudoscalar current:
\be
 \label{ipc}
 P^0=\bar{q}i\gamma_5 q=\bar{u}i\gamma_5 u+\bar{d}i\gamma_5 d
\ee
that can be obtained from the action $\int dy \widetilde{{\cal L}}_{\pi\pi}$
in the same way as before by differentiating with respect to $p^0(x)$ such
that $p(x)=\sum_i p^i(x)\tau^i+ p^0 I$ with $I$ the $2\times 2$ identity
matrix. It is straightforward to check that
\be
 \label{ipcme}
 \langle \Omega|P^0|\pi^k\rangle=\delta^{k3}\frac{8 B c_5}{f}\barro-
 \delta^{k3}\,\frac{8 B^2}{f}\ell_7 \bar{m}
 \equiv \delta^{k3}\widetilde{G}_\pi^\star~,
\ee
where we have explicitly shown the vacuum contribution proportional to
$\ell_7$ from ref.~\cite{gl}. Proceeding along analogous lines as those to
derive eq.\,(\ref{relpa0}), when considering the product $f_t^0
\widetilde{M}_{\pi^0}^2$ one has to add to $\hat{m}\widetilde{G}_\pi$ the
term: 
\be
 \label{ipcrel}
 \bar{m}G_\pi^\star + \bar{m}\frac{8\barro B}{f}c_5~,
\ee
where the first term comes from the vacuum equality $f_\pi M_{\pi^0}^2=
\hat{m}G_\pi+\bar{m}G_\pi^\star$ and the second from the perturbative
expression of the $\pi^0$ mass, eq.\,(\ref{pim}). Collecting both terms and
comparing with eq.\,(\ref{ipcme}) one finally has:
\be
 \label{ipcrel2}
 f_t^0 \widetilde{M}_{\pi^0}^2
 =\hat{m}\widetilde{G}_\pi+\bar{m}\widetilde{G}_\pi^\star~.
\ee
Note that the relations for the charged pion quantities in
eqs.\,(\ref{relpch}) do not get any contribution from $\bar{m}\neq0$ at the
next-to-leading order considered here.

We have also checked that the Ward identity eq.\,(\ref{wi}) is satisfied as
well for non-zero three-momenta, that is, above threshold. This constitutes a
good check for all our former expressions, namely: the spectral relation
between energy and three-momentum, eqs.\,(\ref{spec}), the matrix elements of
the temporal and spatial components of the axial-vector currents with one pion
field, eqs.\,(\ref{caf}), and the pseudoscalar ones eqs.\,(\ref{pc}) and
(\ref{ipcme}).


\section{Three-point Green functions}
\label{sec:anom}
\def\theequation{\arabic{section}.\arabic{equation}}
\setcounter{equation}{0}

In this section we will discuss the in-medium contributions, up to $\Opf$, to
the coupling of pions to a vector source. As will be shown below, the
conservation of the electromagnetic current is nontrivial in this context.
Moreover, the $\rho-$corrections at threshold are of the order of $100\%$ of
the vacuum ones already at $\rho=\rho_0$ due to a large counterterm
contribution proportional to $c_2$.  Nevertheless, for higher energies, the
$\rho-$corrections become considerably smaller compared to the vacuum ones due
to the dominant role of the counterterm $\bar{\ell}_6$ of
SU(2)$_L\times$SU(2)$_R$ CHPT~\cite{gl}. While $c_2$ is saturated by the
$\Delta(1232)$~\cite{bkm1} resonance, $\bar{\ell}_6$ is saturated by the
$\rho(770)$~\cite{gl}. Furthermore, we will show that the surrounding nuclear
medium does not alter the anomalous $\pi^0\rightarrow \gamma\gamma$ decay
amplitude up to $\Opf$. We will also briefly discuss the vanishing of three
pion scattering in the medium.


\subsection{Pion coupling to a photon source}

Let us study the process $\gamma^* \rightarrow \pi^i(\vq) \pi^j(-\vq)$. In the
effective field theory the photon field $A_\mu(x)$ is included via the
external vector source $v_\mu(x)=e A_\mu(x) Q$, with $e\,Q=e\,{\rm diag}(1,0)$
the nucleon charge matrix. The in-medium contributions are depicted in
fig.\,\ref{fig:an3}. In those diagrams where two and more $A^{(2)}$ insertions
are indicated, only one of them has to be considered in the calculation up to
$\Opf$. Indeed the only non-vanishing contribution with an $A^{(2)}$ vertex is
that of fig.\,\ref{fig:an3}a, the other ones are at least $\Ops$. The
calculation is straightforward and for on-shell pions gives:

\begin{figure}[t]
\centerline{\epsfig{file=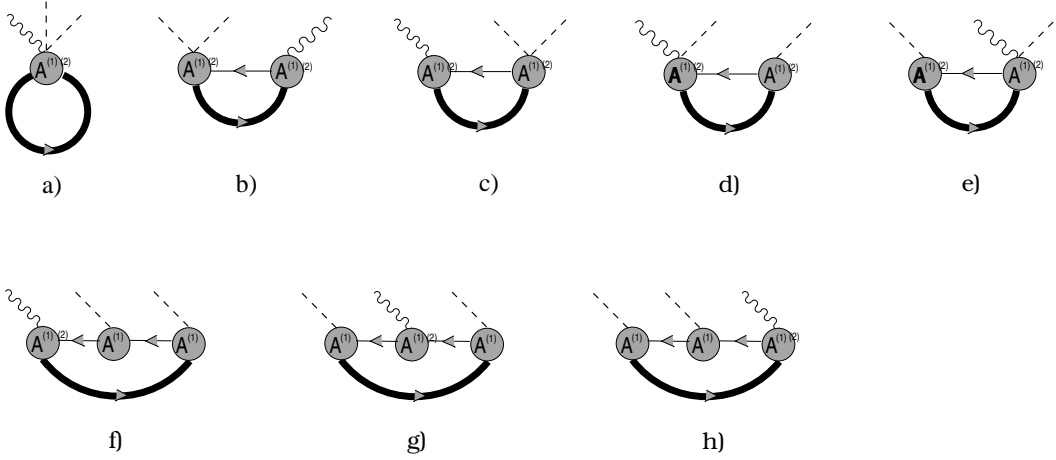,width=5.5in}}
\vspace{0.3cm}
\caption[pilf]{\protect \small 
  In-medium diagrams with two pions and one photon up to $\Opf$. The analogous
  diagrams with more than one Fermi-sea insertion are not shown, as they do
  not contribute here.
\label{fig:an3}}
\end{figure}

\ba
\label{vct}
\langle \pi^+\pi^- |V^0| \Omega\rangle&=&e\left\{ - \frac{\barro}{f^2}+
\frac{g_A^2\vq^2}{\omega^2 f^2}\barro+\omega_+(\vq)-\omega_-(\vq)\right\}~, \\
\label{vcs}
\langle \pi^+\pi^- |V^k| \Omega\rangle&=&-2e \vq^k\,F_V(\vq^2)\left\{
1-\frac{4\hatr}{f^2}c_2-\frac{\hatr+\barro}
 {4m_N f^2}+\frac{g_A^2\barro}{2m_N f^2}+
\frac{g_A^2 (\hatr-\barro)}{4m_N f^2}\frac{\vq^2}{\omega^2}\right\},
\ea
where $F_V(\vq^2)$ is the vacuum pion vector form factor up to $\Opc$ given in
ref.~\cite{gl}. Note that there is no coupling to a $\pi^0\pi^0$ state as
expected from Bose-Einstein statistics.  The term 
$\omega_+(\vq)-\omega_-(\vq)$ in
eq.\,(\ref{vct}) arises because of the difference in the energy of the $\pi^+$
and $\pi^-$ in asymmetric nuclear matter as given by eq.\,(\ref{spec}). Let us
stress that the conservation of the electric current, $\partial_\mu V^\mu=0$,
is nontrivial for the in-medium case, as one can find out by studying the
origin of the different terms in eq.\,(\ref{vct}). The first one comes from
fig.\,\ref{fig:an3}a, the second from figs.\,\ref{fig:an3}f, g and h while the
last two terms originate from from graphs with only two pion lines, without
photons, see eqs.\,(\ref{spec}).  Despite of that, all of them cancel each
other and the temporal derivative is zero as required to guarantee the
conservation of the electric current.  Indeed if one calculates the matrix
element $\langle \pi^+\pi^-|\partial_\mu V^\mu|\Omega \rangle$ from
eq.\,(\ref{vct}) and (\ref{vcs}), one obtains: 
\ba
 \label{edif}
 \langle \pi^+\pi^-|\partial_\mu V^\mu|\Omega \rangle
 &\propto& 
 2\omega (\vq)\left(
 -\frac{\barro}{f^2}+\frac{g_A^2\vq^2}{\omega^2 f^2}\barro\right)
 +\omega_+(\vq)^2-\omega_-(\vq)^2
 \nn\\
 &=&
  2\omega(\vq)\left(-\frac{\barro}{f^2}+
 \frac{g_A^2\vq^2}{\omega^2 f^2}\barro\right)+\left(2\frac{\barro}{f^2}
 \omega (\vq)-
 \frac{2g_A^2\barro \vq^2}{f^2 \omega(\vq)}\right)\nn\\
 &=&0~.
\ea

\noindent
The in-medium contributions to the spatial components of the vector current,
eq.\,(\ref{vcs}), come from figs.\,\ref{fig:an3}a for the first
density-dependent term inside the curly brackets, from \ref{fig:an3}b, c for
the second one, and from fig.\,\ref{fig:an3}f, g, h for the last two terms.
Since it does not involve the non-relativistic suppression factor $1/m_N$, the
dominant term is the one with the rather large $c_2$ counterterm which is also
enhanced by a factor of 2 as compared, e.g., to eq.\,(\ref{pv}).  The third
term in eq.\,(\ref{vcs}), figs.\,\ref{fig:an3}b and c, involves only the
proton density since only protons couple to the photon at this order. The
contribution from figs.\,\ref{fig:an3}d and e are canceled by those of the
wave function renormalization. The last term comes from fig.\,\ref{fig:an3}g
and this is why it only involves the neutron density. The term before the last
one is a combination of figs.\,\ref{fig:an3}f and h, with only proton density
dependence, and that of fig.\,\ref{fig:an3}g.

\begin{figure}[htb]
\centerline{\epsfig{file=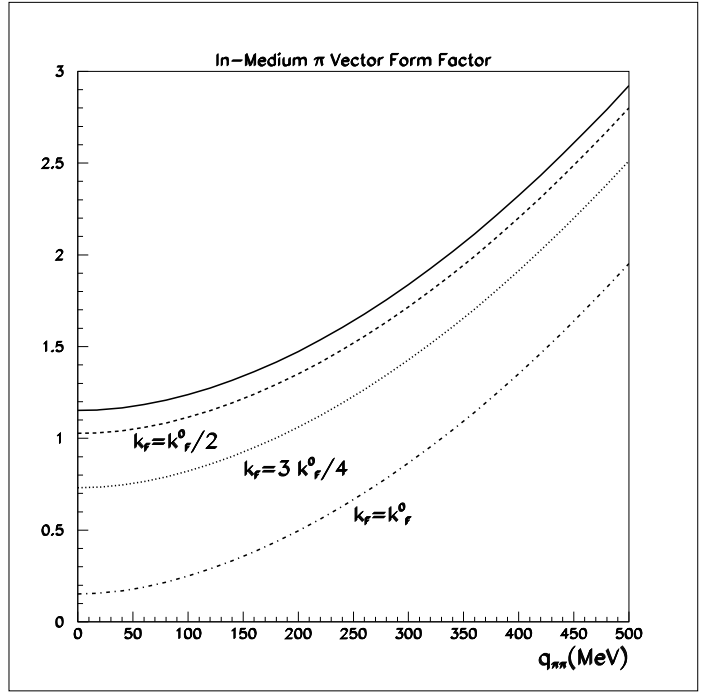,width=3.6in}}
\vspace{0.3cm}
\caption[pilf]{\protect \small In-medium pion vector form factor in 
  symmetric nuclear matter as function of the modulus of the pion 3-momentum
  $q_{\pi\pi}$ (= $|{\bf q}|$). The continuum line is the vacuum $\Opc$ result
  from ref.~\cite{gl}. The dashed, dotted and dashed-dotted lines are the
  in-medium results up to $\Opf$ and correspond to $k_F=k_F^0/2$, $3k_F^0/4$
  and $k_F^0$, respectively. Note that at $q_{\pi\pi}=0$ the photon virtuality
  is four times the pion mass squared and that the {\em vacuum} form factor is
  one at virtuality zero.
\label{fig:ffv}}
\end{figure} 

\noindent
In fig.\,\ref{fig:ffv} we show the in-medium results to the spatial vector
coupling in symmetric nuclear matter for three values of $k_F$: $k_F^0/2$, $3
k^0_F/4$ and $k_F^0$. As stated above, the modifications with density increase
rather fast due to the large $c_2$ counterterm. Note, however, that the
relativistic corrections become more important for higher three-momenta. In
fact, the difference between the vacuum $\Opc$ result, solid line, with
respect to any of the other lines, gives the in-medium contributions.  They
rapidly increase as soon as we approach $k_F^0$ because of the cubic
dependence on $k_F$ over a quantity of order $k_F^0$.  The lowest order $\Opd$
result is just $1$. Taking the resonance saturation of the $c_2$ into account,
see ref.~\cite{bkm1}, one can see that this counterterm just reflects the
impact of the $\Delta(1232)$ in low energy $\pi N$ scattering. Thus the term
proportional to $c_2$ in eq.\,(\ref{vcs}), stemming from fig.\,\ref{fig:an3}a,
establishes a large $\Delta$-hole contribution. Nevertheless, for larger
values of the three-momentum $|\vq|$, the in-medium corrections are relatively
less important than the vacuum ones, due to the large $\bar{\ell}_6$ CHPT
counterterm that signals the well-known and very important $\rho(770)$
contribution to the pion vector form factor in vacuum \cite{gl}.
Finally note that the form factor in nuclear matter does not go to one at zero
virtuality, because this background carries charge.

\subsection{Anomalous sector}
\label{an}

Since the effective Lagrangian $\widetilde{{\cal L}}_{\pi\pi}$ contains terms
with odd number of pions including a $\gamma_5$ matrix, one might expect the
in-medium appearance of processes violating intrinsic parity, as, e.g., the
scattering of three pions or contributions to the $\pi^0\rightarrow \gamma
\gamma$ decay amplitude.  We will show that, up to $\Opf$, there are no
in-medium corrections to the $\pi^0\rightarrow \gamma \gamma$ amplitude and
that the three-pion scattering vanishes in unpolarized nuclear matter at the
same order as well.

\begin{figure}[htb]
\centerline{\epsfig{file=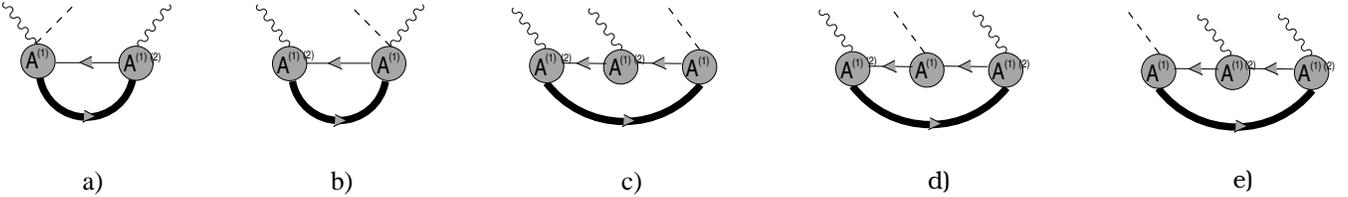,width=7.in}}
\vspace{0.3cm}
\caption[pilf]{\protect \small 
  In-medium diagrams with two photon and one pion lines 
  up to $\Opf$. The analogous diagrams with more than one Fermi-sea insertion
  or with only one $A$ vertex are not shown. Every diagram vanishes
  independently up to $\Opf$.
\label{fig:an1}}
\end{figure}

The set of diagrams to be considered are shown in fig.\,\ref{fig:an1}. The
figure does not contain any diagram with only one $A$ vertex, as such a
contribution is trivially zero. Namely, when $A=A^{(1)}$, it vanishes because
then we have one $\gamma_5$ matrix together with, at most, two other $\gamma$
matrices, such that the Dirac trace is zero. When $A=A^{(2)}$, the result is
zero, simply for the reason that there are no terms in $A^{(2)}$ with two
photons and one pion.  As indicated in the diagrams of fig.\,\ref{fig:an1},
each diagram can contain an $A^{(2)}$ operator attached to a photon line. In
fact, one could also have diagrams with just two photon lines stemming from
the same $A^{(2)}$ vertex; but they vanish since $[v_\mu(x),v_\nu(x)]=0$.

Let us show that all the diagrams of fig.\,\ref{fig:an1} vanish.  Consider
first the diagrams fig.\,\ref{fig:an1}a and b with only two leading $A^{(1)}$
operators.  It is easy to prove that the flavor trace vanishes for each of
them. For this case $\Gamma_\mu(x)=-i v_\mu(x)$ and
$\Delta_\mu(x)=-[\phi(x),v_\mu(x)]/2$.  As a result the flavor traces in
figs.\,\ref{fig:an1}a,b vanish since: 
\ba
\label{tracf1}
  \hat{n}(p)\langle \Gamma_\mu(x)\Delta_\nu(y) \rangle 
 &\rightarrow& \langle Q[\phi,Q]\rangle=0~,
\nn\\
\bar{n}(p)\langle \tau_3 \Gamma_\mu(x)\Delta_\nu(y) \rangle 
 &\rightarrow& 
\langle \tau_3 Q[\phi,Q]\rangle=0~,
\ea
where $Q$ is the charge matrix.  A change of the order of $\Gamma_\mu(x)$ and
$\Delta_\nu(y)$ in the equations above does not modify the results since
$[\tau_3,Q]=0$. For the same diagrams, but now with one $A^{(2)}$ operator
replacing one $A^{(1)}$ operator, there is only a contribution when the other
$A^{(1)}=-ig_A \gamma^\mu \gamma_5 \Delta_\mu$, with $\Delta_\mu$ as given
above.\footnote{The simplest way to realize this is by considering the
  $\varphi-$counting introduced in ref.~\cite{gl} where $\phi \sim \varphi$,
  $a_\mu \sim \varphi$, $p \sim \varphi$, $v_\mu \sim \varphi^2$ and $s\sim
  \varphi^2$. In this counting, the $\Delta_\mu$ operator always gives rise to
  terms with odd powers of $\varphi$ and the $\Gamma_\mu$ operator as well as
  $A^{(2)}$ to even power terms.} For the terms in $A^{(2)}$ without gamma
matrices the result vanishes, after taking the Dirac trace, because of the
$\gamma_5$ accompanying $\Delta_\mu$ in $A^{(1)}$ and the fact that only three
extra gamma matrices are available. The contribution proportional to $c_4$ in
$A^{(2)}$ vanishes because there are no terms in $u_\mu$ involving vector
sources without pion fields. Hence two or more pions would arise and this is
not allowed. The terms with $c_6$ and $c_7$ involve the $F^+_{\mu\nu}$, which
for our present purposes simplifies to $2\partial_\mu v_\nu-2\partial_\nu
v_\mu$. Thus they only contain the charge matrix $Q$ in flavor space. In
summary, when $\Gamma_\mu$ is replaced by $F^+_{\mu\nu}$, the same reasoning
as previously given for the corresponding diagrams with only $A^{(1)}$
operators holds also for this case.  The diagrams fig.\,\ref{fig:an1}c, d and
e deserve special attention, although in the end each of them also vanishes
independently at $\Opf$ because of the Dirac trace after symmetrization with
respect to the photon lines. Furthermore it is easy to see that the diagrams
fig.\,\ref{fig:an1}c, d, and e with one $A^{(2)}$ vertex, attached to a
$v_\mu(x)$ external source, vanish as well. Since these diagrams are at least
$\Opf$, one can replace the momentum $p_j$ by $p$ inside the expression
$\barr{p}_j+m_N$, which enters in the numerator of the baryon propagator,
because $p_j=p+Q_j$ and the difference will be $\Ops$. Furthermore, for the
same reason, we can set $\barr{p}$ equal to $\gamma_0 m_N$. Taking also into
account that the four-momentum $Q$ of the $\pi^0$ is just $(M_\pi^0,\bf{0})$
it is straightforward to see that in all the Dirac traces one has the product
$\gamma_0(I+\gamma_0)\gamma_5 (I+\gamma_0)=0$ which implies the vanishing of
the $\Opf$ contributions from such diagrams.

As a result of this discussion, taking also into account that there are no
contributions from diagrams with two or three Fermi-sea insertions because of
the reasons discussed in sec.~\ref{sec:prop}, the final in-medium
$\pi^0\rightarrow \gamma\gamma$ decay amplitude
$\widetilde{T}(\pi^0\rightarrow \gamma\gamma)$, up to $\Opf$, just corresponds
to the pure vacuum one given by the Wess-Zumino-Witten term in the effective
chiral action \cite{an}, counted as $\Opc$ in CHPT. Note also that the
corrections to the vacuum phase space induced by the change of the $\pi^0$
mass in the medium are at least ${\cal O}(p^{9})$, of the same order as the
neglected contributions to the decay resulting from $\Ops$ contributions to
$\widetilde{T}(\pi^0\rightarrow \gamma\gamma$). Thus up the accuracy we are
considering in this paper, there is no modification of the $\pi^0\rightarrow
\gamma\gamma$ width due to the nuclear medium.

Finally, let us consider the 3$\pi$ scattering, which is not allowed in
vacuum, since the anomalous WZW terms only give rise to processes with five or
more Goldstone bosons. It turns out that the $3\pi$ scattering does not occur
in nuclear matter, at $\Opf$, either\footnote{These will be subprocesses in
  the discussion of the more realistic $4\pi$ scattering in the next
  section.}.
\begin{figure}[htb]
\centerline{\epsfig{file=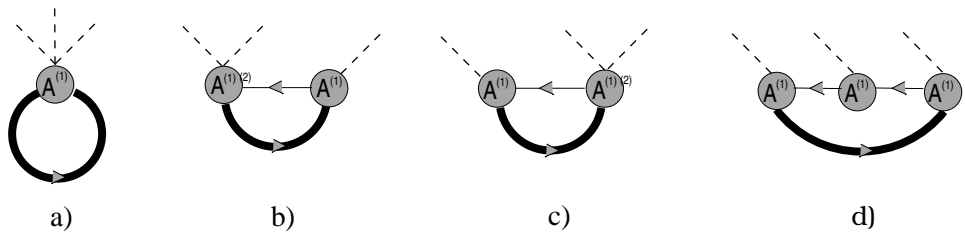,width=5.in}}
\vspace{0.3cm}
\caption[pilf]{\protect \small 
  In-medium diagrams with $3\pi$ lines up to $\Opf$. 
  Each of them vanishes independently in asymmetric nuclear matter. The
  analogous diagrams with more than one Fermi-sea insertion are not shown.
\label{fig:an2}}
\end{figure}
The set of diagrams to be evaluated is depicted in fig.\,\ref{fig:an2}.  The
diagram of fig.\,\ref{fig:an2}a vanishes trivially because of the Dirac trace
since we have one $\gamma_5$ and at most two gamma matrices while at least
four gamma matrices are necessary. The latter structure is possible for
diagrams fig.\,\ref{fig:an2}b,c,d; but in this case one is left with the
totally antisymmetric tensor in four dimensions,
$\epsilon^{\mu\nu\rho\sigma}$, which requires four independent momenta.
However, we have only three of them in any diagram: one is the in-medium
running on-shell four-momentum $p$ together with two four-momenta from the
pions (the third one is given by energy-momentum conservation). Thus these
contributions vanish. Note that only diagrams with just one Fermi-sea
insertion are shown in fig.\,\ref{fig:an2}, since the proof that each diagram
vanishes independently is only based on the Dirac structure of the diagram
which is the same independently of whether a free baryon propagator is
replaced by a Fermi-sea insertion or not.


\section{Four-point Green functions: Pion scattering}
\label{sec:psc}
\def\theequation{\arabic{section}.\arabic{equation}}
\setcounter{equation}{0}

\begin{figure}[htb]
\centerline{\epsfig{file=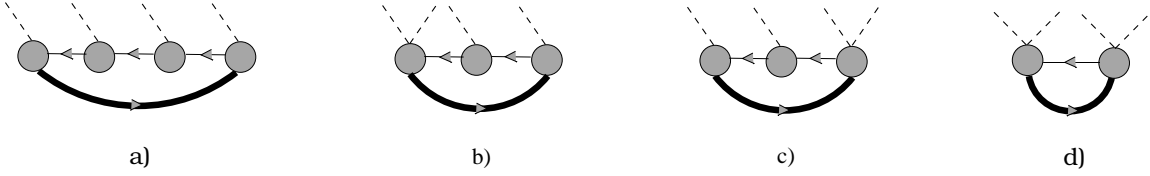,width=6.in}}
\vspace{0.3cm}
\caption[pilf]{\protect \small 
  In-medium diagrams contributing to $4\pi$ scattering 
  at $\Opt$. All the vertices are $A^{(1)}$ operators.
\label{fig:scap1}}
\end{figure}

The in-medium $4\pi$ scattering contributions begin to appear at $\Opt$
because an in-going pion four-momenta can be combined with an out-going one,
such that $Q_j^0\approx 0$ for one of the propagators. The corresponding
diagrams are shown in fig.~\ref{fig:scap1}.  One can see the $\Opt$ behavior
just by applying eq.\,(\ref{ffc}), case b). If all the four external pion legs
had $Q^0\sim k_F^2/2m_N$, then the chiral power counting would start at
$\nu=1$ with $V_\rho=1$. However, since the pion legs are of course on-shell
with $Q^0\geq M_\pi$, fig.\,\ref{fig:scap1}a starts at $\Opt$. Indeed, the
analysis of all the possible diagrams according to eq.\,(\ref{ffc})-b) up to
$\Opt$ shows that the leading contribution corresponding to the diagrams shown
in fig.\,\ref{fig:scap1} appears at $\nu=3$.

Only one Fermi-sea insertion has to be considered in fig.\,\ref{fig:scap1}.
There is no way to include three or four of them, because there is always one
four-momentum $Q^0\sim M_\pi$ running along one insertion that makes it
impossible to satisfy $p_j^2=m_N^2$ for this thick line, as discussed in
sec.~\ref{sec:prop}. However, it is in principle possible to have two
Fermi-sea insertions.  Nevertheless, as shown in sec.~\ref{sec:prop}, in this
case there is a cancellation, below the Fermi-momentum, between the diagrams
with two Fermi-sea insertions and the imaginary parts of those with only one
Fermi-sea insertion. It is easy to see that for $Q^0=0$ both contributions
cancel each other: setting $p_j=p+Q$ and then imposing $p_j^2=m_N^2$, one can
easily see that:
\be
\label{c1}
{\mathbf Q}^2+2{\mathbf{p}}{\mathbf{Q}}=0 ~,
\ee
since $Q^0=0$ and $p^2=m_N^2$. On the other hand, because of the previous
cancellations, $|{\mathbf{p}}+{\mathbf{Q}}|> k_F$. By squaring this condition
and imposing eq.\,(\ref{c1}), one concludes ${\mathbf p}^2>k_F^2$ that is not
allowed.

\begin{figure}[htb]
\centerline{\epsfig{file=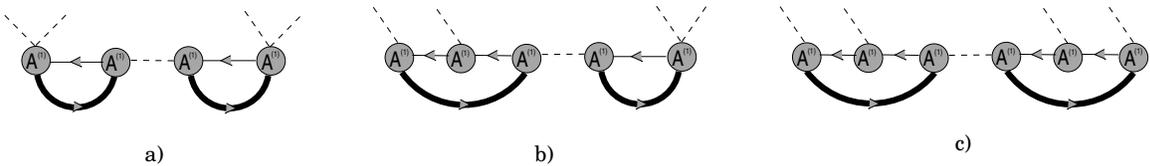,width=6.in}}
\vspace{0.3cm}
\caption[pilf]{\protect \small 
   In medium-pion diagrams contributing at $\Opc$. Each 
  of them vanishes independently. More details are given in the text.
\label{fig:scap2}}
\end{figure}

In fig.\,\ref{fig:scap2} we show some $\Opc$ diagrams that are two connected
$3\pi$ scattering processes, which vanish, as already discussed in the
previous section, because of the appearance of the totally antisymmetric
tensor $\epsilon^{\mu\nu\sigma\rho}$.  Nevertheless the vanishing of the
diagrams in fig.\,\ref{fig:scap2} is due to the pseudoscalar character of the
exchanged pion.  If instead there were short range NN forces, these diagrams
would not only have survived, but would be enhanced by the appearance of
nucleon mass factors from each bubble, as discussed in sec.~\ref{breakdown}.

In the following we will restrict ourselves to the case of symmetric nuclear
matter $\barro=0$. The results presented here have been calculated in two
ways: using the relativistic formalism from the beginning and performing the
chiral expansion up to $\Opc$ afterwards {\em or}, from the beginning, making
use of those simplifications that are allowed order by order. The results are
the same under both ways.  Because of the loss of crossing symmetry due to the
absence of Lorentz covariance, we consider separately the scattering
amplitudes:
\ba
\label{scatam}
T_1 \equiv \pi^0\,(q_1)\pi^0\,(q_2) &\!\!\rightarrow\!\!& 
 \pi^0\,(q_3)\pi^0\,(q_4)~, \nn\\
T_2 \equiv \pi^+(q_1)\pi^+(q_2)  
 &\!\!\rightarrow\!\!&\pi^+(q_3)\pi^+(q_4) ~,\nn\\
T_3 \equiv \pi^+(q_1)\pi^-(q_2)  
 &\!\!\rightarrow\!\!&\pi^+(q_3)\pi^-(q_4)~,
\ea

where $q_i$ is the four-momentum of each pion with
$\vq_1+\vq_2=\vq_3+\vq_4=0$.  We further define the three-momenta ${\mathbf
  Q}={\mathbf q}_1-{\mathbf q}_3$ and ${\mathbf R}={\mathbf q}_1+{\mathbf
  q}_3$ together with the usual Mandelstam variables
$s=(q_1+q_2)^2=(q_3+q_4)^2$, $t=(q_1-q_3)^2=(q_2-q_4)^2=-{\mathbf Q}^2$ and
$u=(q_1-q_4)^2=(q_2-q_3)^2= -{\mathbf R}^2$.  It is worthwhile to mention that
the $\Opc$ corrections to the leading $\Opt$ diagrams in fig.\,\ref{fig:scap1}
vanish. Note as well that the in-medium corrections to the pion mass in the
symmetric nuclear matter, eq.\,(\ref{pim}), as well as the wave function
renormalization, eq.\,(\ref{wfr2}), appear first at $\Opf$.  Then up to $\Opt$
and in the $\rho-$isospin limit $\bar{m}=\barro=0$, one has the final
expressions: 
\ba
 \label{fex}
 T_1(\pi^0\,\,\pi^0 \rightarrow \pi^0\, \,\pi^0 )
 &=&
 \frac{M_\pi^2}{f_\pi^2}+\frac{g_A^4 m_N t u}
 {f^4 \omega^2}\left[\ci(0,{\mathbf Q})+
 \ci(0,{\mathbf R})\right] ~,\nn\\
 T_2(\pi^+ \pi^+ \rightarrow \pi^+ \pi^+)
&=&
 \frac{2M_\pi^2-s}{f_\pi^2}-\frac{4\omega^2m_N}{f^4}\left[\ci(0,{\mathbf Q})+
 \ci(0,{\mathbf R})\right]+\frac{4g_A^2 m_N}{f^4}
 \Big\{\left(t+2\omega^2-2M_\pi^2\right)
 \ci(0,{\mathbf Q})\nn\\
&+&
 \left(u+2\omega^2-2M_\pi^2\right)\ci(0,{\mathbf R})\Big\}
 +\frac{2g_A^4m_N}{f^4\omega^2}\left\{tu-2(\omega^2-M_\pi^2)^2\right\}
 \left[\ci(0,{\mathbf Q})+
 \ci(0,{\mathbf R})\right]~,\nn\\
 T_3( \pi^+ \pi^- \rightarrow \pi^+ \pi^-)&=&\frac{2M_\pi^2-u}{f_\pi^2}+
 \frac{4\omega^2 m_N}{f^4}\ci(0,{\mathbf Q})-
 \frac{4g_A^2 m_N}{f^4}
 \left\{t+2\omega^2-2M_\pi^2\right\}\ci(0,{\mathbf Q})\nn\\
&+&
 \frac{4g_A^4 m_N}{f^4 \omega^2}(\omega^2-M_\pi^2)^2\ci(0,{\mathbf Q})~,
\ea
with $\omega$ the energy of any of the pions and the function $\ci(Q^0,{\mathbf
  {Q}})$ as defined in eq.\,(\ref{funi}). Notice the presence of the $2 m_N$
factor accompanying the $\ci(0,{\mathbf {Q}})$ and $\ci(0,{\mathbf {R}})$
functions that, as discussed in sec.~\ref{breakdown}, implies a dramatic
reduction of the chiral scale $\Lambda$ to about 170 MeV.
Nevertheless, one should keep in mind that this scale 
only refers to the Fermi-momentum $k_F$ and the
three-momenta $\bf{Q}$ and $\bf{R}$, but not to the energy $\omega$ of the
pions which is bounded from below by twice the pion mass.
The contributions in eqs.\,(\ref{fex}) proportional to $g_A^4$ and $g_A^2$
come from fig.\,\ref{fig:scap1}a and fig.\,\ref{fig:scap1}b,c, respectively.
Those stemming from fig.\,\ref{fig:scap1}d originate from the
Weinberg-Tomozawa term and do not contain any $g_A$ factor.

\begin{figure}[htb]
\centerline{\epsfig{file=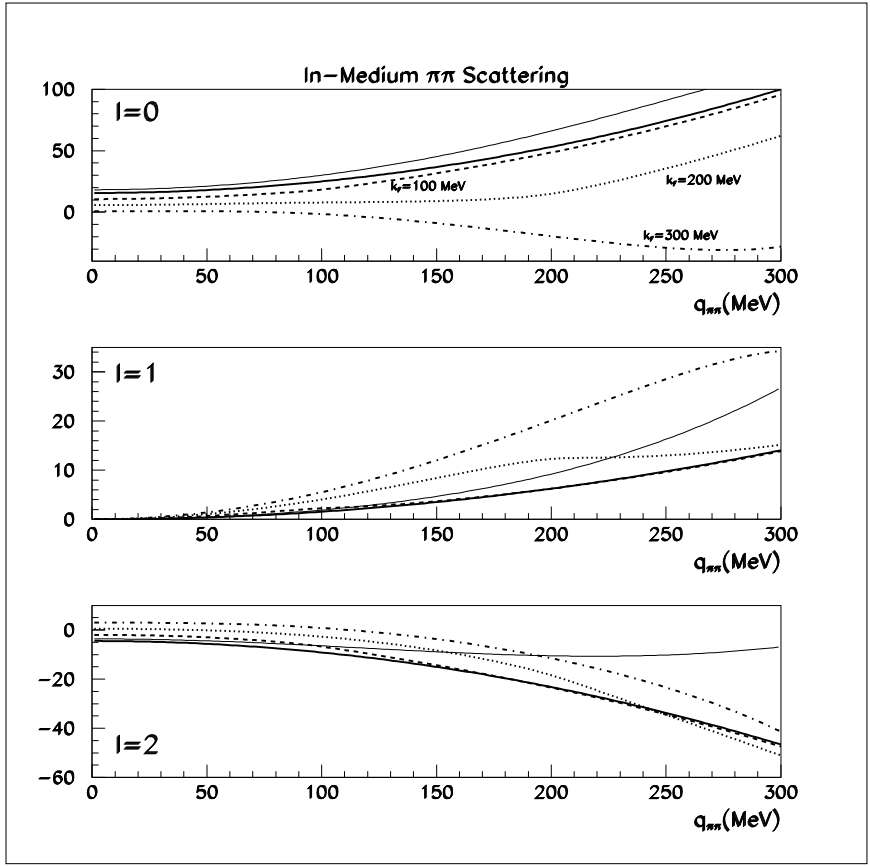,width=4.5in}}
\vspace{0.3cm}
\caption[pilf]{\protect \small {}From top to bottom: S-wave I=0, 
  P-wave I=1 and S-wave I=2 $\pi\pi$ partial wave amplitudes. The thick full
  lines are the leading $\Opd$ $\pi\pi$ scattering amplitudes while the thin
  solid ones are the pure vacuum $\Opc$ results from ref.~\cite{gl}. The
  dashed, dotted and dashed-dotted lines are the $\Opt$ in-medium partial wave
  amplitudes for $k_F=100$, 200 and 300 MeV, respectively.
\label{fig:scatps}}
\end{figure}

{}From the previous expressions one can easily work out the in-medium
amplitudes with well defined isospin (I) and angular momentum. These are shown
in figs.\,\ref{fig:scatps} where, from top to bottom, the first panel
corresponds to the S-wave I=0 amplitude, the second one to the P-wave I=1
partial wave and the third panel to the S-wave I=2 amplitude, for three values
of the Fermi-momentum $k_F$ as indicated in the figure. The thick solid curves
correspond to the lowest order $\pi\pi$ CHPT partial waves while the thin
solid ones to the pure vacuum $\Opc$ $\pi\pi$ amplitudes, ref.~\cite{gl}.  One
can see strong in-medium corrections close to threshold for the I=0 and 2
S-waves induced by the diagram of fig.\,\ref{fig:scap1}d.  This diagram only
contains the Weinberg-Tomozawa term proportional to $\Gamma_\mu$,
eq.\,(\ref{a1}), which does not involve a pion three-momenta as is the case
for the term proportional to $g_A$ in eq.\,(\ref{a1}). This diagram has so
far not been considered in the works (see \cite{vacas,mona,dj}) that are
involved in the physics of in-medium pion scattering, triggered by the
experimental data of the CHAOS Collaboration \cite{benutti}. The latter
reported originally a dramatic enhancement of the $(\pi^+,\pi^+\pi^-)$
reaction close to the $2\pi$ threshold, although opposite conclusions have
recently been obtained by the Crystal Ball Collaboration \cite{bnl} from the
study of the reaction $(\pi^-,\pi^0\pi^0)$.  It is nowadays believed that when
one properly accounts for the limited acceptance of the CHAOS detector, these
results are mutually consistent~\footnote{See the last entry of
  ref.\cite{benutti}.  However, this issue is still
  controversial~\cite{JoeComford}.}.  As one can see, the diagram
fig.\,\ref{fig:scap1}d is indeed dominant for small pion three-momenta, as far
as this perturbative analysis holds, since it is not kinematically suppressed
as the other contributions with $g_A$ dependence because of the P-wave
coupling of the pions to the nucleons.  At this point it is also worthwhile to
note that we have shown explicitly, eq.\,(\ref{fex}), that the in-medium
contribution in symmetric nuclear matter to $\pi\pi$ scattering from the
leading pion-nucleon Lagrangian, ${\cal L}^{(1)}_{\pi N}$, does not vanish as
stated in ref.~\cite{dj}.  On the other hand, our results for the S-waves,
valid for nuclear matter, cannot be taken literally for finite nuclei when
$Q^0=0$, ${\bf Q}=0$ since the ${\cal I}(Q^0,{\bf Q})$ function should then
vanish due to the finite excitation energy of a few MeV of the nuclear energy
levels. This effect is pointed out in ref.~\cite{navarro}. This would imply
that in finite nuclei the S-wave I=0 and I=2 amplitudes at threshold do not
obtain any contribution from fig.\,\ref{fig:scap1}d, although, as soon as the
pion three-momentum is different from zero, these large corrections rapidly
build up. However, as energy increases, the diagrams of fig.\,\ref{fig:scap1}b
and c, which were as well not considered in ref.~\cite{vacas,mona,dj}, become
more relevant, dominating the in-medium corrections to the P-wave I=1
scattering amplitude. On the other hand, their influence is rather small for
the S-waves. The term proportional to $g_A^4$, fig.\,\ref{fig:scap1}a, absent
as well in refs.~\cite{vacas,mona,dj}, is very relevant for the S-waves when
the pion three-momentum is higher than 100 MeV, although its influence is much
smaller for the P-wave.  It is remarkable to point out that for the I=0 and
I=1 cases the amplitudes change dramatically with density, while for the I=2
case the S-wave scattering amplitude receives in-medium corrections smaller
than those from vacuum $\Opc$ CHPT when $|\vq|>200$ MeV.  Moreover, the I=0
partial waves changes even sign, becoming repulsive, while in the I=1 case
there is a strong enhancement as soon as one moves to higher three-momenta and
Fermi-momentum. We remind the reader that in the present situation with
$Q^0_j\sim 0$ the perturbative results are only valid for $k_F$ and
three-momenta up to around $170$ MeV and that the dotted and dashed-dotted
lines in fig.\,\ref{fig:scatps} are just shown to point out the huge in-medium
corrections as soon as one goes beyond that limit, although in the end the
actual size of the in-medium corrections depend on the specific reaction.
Indeed, for the physics relevant to the CHAOS experiment the effective average
density is around $\rho_0/2$ \cite{vacas} corresponding to a $k_F\sim 210$
MeV. Hence, one would expect large in-medium non-perturbative effects to take
place.  Nevertheless, the different models dealing with this problem should
match at low densities and momenta with the $\Opt$ results presented in
eq.\,(\ref{fex}) in order to take care of the requirements of chiral symmetry.

\section{Conclusions}
\label{sec:conclusions}

In this paper we have tackled the problem of establishing a chiral effective
field theory in nuclear matter with explicit pion fields. We have made use of
the results of ref.~\cite{med1} where, by integrating out the baryonic fields
in the path integral representation of the generating functional, once the
change of ground state from vacuum to nuclear matter is done, the in-medium
chiral SU(2)$\times$SU(2) Lagrangian is derived. We have briefly reviewed
these results in appendix~\ref{subsec:genfunc} as well as the perturbative
case presenting the corresponding Feynman rules. After establishing the power
counting rules, we have systematically studied several low--energy QCD Green
functions up to next-to-leading order, $\Opf$, when the standard counting
holds or by working out the leading in-medium contributions in the
non-standard case.  The novel results obtained here can be summarized as
follows:
\begin{itemize}
\item[(1)] In contrast to previous works, which apply the mean-field approach
  or many--body calculations, the in-medium chiral counting is worked out in
  sec.~\ref{sec:counting}.  The counting scheme is dependent on the energy
  flowing into the nucleon lines. This leads one to consider the standard and
  the non--standard case, respectively, as summarized in
  table~\ref{tab:count}. In the former case, the chiral expansion of pion
  properties in the medium starts with terms at ${\cal O}(p^4)$, and the
  next--to--leading order corrections appear at ${\cal O}(p^5)$, quite
  different to the in--vacuum power counting.
  \renewcommand{\arraystretch}{1.2}
\begin{table}[H]
\begin{center}
\begin{tabular}{|c|cc|}
\hline
&   Standard counting & Non-standard counting \\
\hline
energy flow  & $Q^0 \sim M_\pi \sim \Op$ & $Q^0 \sim Q^2/2m_N \sim \Opd$ \\
nucleon propagator & $D_0^{-1} \sim {\cal O}(p^{-1})$ & $D_0^{-1} \sim
{\cal O}(p^{-2})$\\
counting index $\nu$ & eq.\,(\ref{fc}) & eq.\,(\ref{ffc}) \\
breakdown scale $\Lambda$ & $\sqrt{6}\pi f_\pi \simeq 0.7\,$GeV &
$6\pi^2 f_\pi^2/2m_N \simeq 0.27\,$GeV \\
\hline
\end{tabular}
\end{center}
\vspace{-0.2cm}
\centerline{\parbox{12cm}{
\caption{Chiral counting for in--medium CHPT for the standard and the
non--standard case. The pertinent breakdown scales are also given.
Here, ``energy flow'' means the energy flowing into the nucleon lines
mediated by pions or external sources (as defined in sec.~\ref{sec:counting}).
\label{tab:count}}}}
\end{table}

\item[(2)] We have also established the relevant scales of the problem when
  restricting ourselves to the $\La_{\pi\pi}$ and $\La_{\pi N}$ Lagrangians,
  see sec.~\ref{sec:gf} or ref.~\cite{med1}. In the vacuum, the pertinent
  scale is $\Lambda_\chi \simeq 1$ GeV$\sim 4\pi f_\pi$.  In the medium, one
  has two new scales. These are: $\sqrt{6} \pi f_\pi \simeq 0.7$ GeV and
  $6\pi^2 f_\pi^2/2m_N\simeq 0.27$ GeV, in case that the standard or the
  non-standard counting rules apply, see also table~\ref{tab:count}. We point
  out that in case of P-wave interactions, these scales are reduced by factors
  $1/g_A$ and $1/g_A^2$, respectively.
 
\item[(3)] We have studied the quark condensates and re-derived, from the
  effective field theory point of view, known results in symmetric nuclear
  matter, and have further extended them to the non-symmetric case.
  
\item[(4)] We have considered the propagation of pions in the medium obtaining
  the pion propagator up to $\Opf$. In this way we have established that
  chiral symmetry can account for the observed shift of the mass of the
  negative pion in deeply bound pionic states in $^{207}$Pb. Our numerical
  result $\Delta M_{\pi^-} = 18\pm 5\,$MeV is compatible with the experimental
  number, $\Delta M_{\pi^-} = 23 - 27\,$MeV \cite{g,i} within errors.
  
\item[(5)] The wave function renormalization of the pion fields corresponding
  to the calculated in-medium action $\int dx \widetilde{\La}$ has been
  established making use of first principles.
  
\item[(6)] We have also studied the coupling of pions with axial-vector and
  pseudoscalar sources. In particular, it is shown that in-medium corrections
  up to $\Opf$ do not spoil the validity of the Gell-Mann-Oakes-Renner
  relation.  We have also found a decrease with increasing density for both
  the quark condensates and the temporal component of the pion decay constant
  $f_t$. Both effects seem to indicate a partial chiral symmetry restoration
  with increasing density. We remark again that a systematic study of the
  in-medium order parameters still has to be performed.  A drastic quenching
  with density has also been obtained for the spatial component of the pion
  decay constant $f_s$.  To $\Opf$ we have checked the QCD Ward identity
  relating both the temporal and spatial components of the axial-vector
  currents with the pseudoscalar ones and quark masses.
  
\item[(7)] A rapid decrease with density of the coupling of a photon to two
  pions, particularly in the threshold region, has been found. The derived
  vector current amplitudes coupled to two pions fulfill the requirement of
  current conservation. Furthermore, we have established the absence of
  in-medium renormalization up to $\Opf$ of the anomalous $\pi^0\rightarrow
  \gamma\gamma$ decay amplitude.
  
\item[(8)] Finally, $\pi\pi$ scattering has been studied up to $\Opt$ since in
  this case the non-standard counting occurs. As explained in
  secs.~\ref{sec:counting} and \ref{sec:psc}, this implies that the in-medium
  corrections start at lower orders than in the standard case, here already at
  $\Opt$. In addition the scale, below which the perturbative expansion is
  applicable, decreases.  As a result the in-medium corrections increase very
  rapidly with density and already at $k_F\simeq 200$ MeV, or at a density of
  just $\sim 0.4\rho_0$, they are $100\%$ with respect to the lowest order
  CHPT results. The diagrams presented in figs.\,\ref{fig:scap1} are the
  leading ones in the chiral expansion and have not been considered so far in
  the literature.
\end{itemize}

\bigskip

Future challenges are the inclusion of multi-nucleon contact interactions,
which are enhanced because of the largeness of the S-wave scattering lengths
related to the presence of shallow NN bound states, as well as the
simultaneous calculation of pion loops necessary for determination of the full
$\Ops$ contributions. Furthermore, the possibility of some non-perturbative
scheme that allows for the recovery of the scale $\sqrt{6}\pi f_\pi$, even in
the case of the non-standard counting or when the multi-nucleon local
interactions are included, should be pursued.

\newpage

\noindent {\bf Acknowledgments}

\medskip We would like to thank E. Oset for useful discussions.  The work of
J.A.O. was supported in part by funds from DGICYT under contract PB96-0753 and
from the EU TMR network Eurodaphne, contract no.  ERBFMRX-CT98-0169.


\begin{appendix}
\section{Generating functional}
\label{subsec:genfunc}
\def\theequation{\Alph{section}.\arabic{equation}}
\setcounter{equation}{0}  

Our starting point is the generating functional $\cZ(v,a,s,p)$ in the presence
of external vector $v$, axial--vector $a$, scalar $s$ and pseudoscalar $p$
sources (included in $\La_{\pi\pi}$ and $\La_{\pi N}$) which is obtained after
integrating out the baryonic fields, as derived in ref.~\cite{med1} (for all
details on the derivation, we refer to that paper). It is given by 
\ba
 \label{fZ2}
 e^{i \cZ[v,a,s,p]}\!\!\!&=&\!\!\!\int\! 
 [dU] \exp\Bigg\{i\!\int\! dx\, \La_{\pi\pi}\!
\nonumber \\ &-& i 
\!\int\!\frac{d\vp}{(2\pi)^3 2 E(p)} \!\int\! dx\,dy\, e^{ip(x-y)}\,
\hbox{Tr}\Bigg( A[I_4-D_0^{-1}A]^{-1}
\arrowvert_{(x,y)}
(\barr{p+m_N})\,n(p)\Bigg)
\nonumber \\ 
&+&\frac{1}{2}\!\int\!\!\! \frac{d\vp}{(2\pi)^3 2E(p)}\!\int\! \!\! 
\frac{d\vq}{(2\pi)^3 2 E(q)}
\!\int\! dx\,dx'\,dy\,dy'\,e^{ip(x-y)}
e^{-iq(x'-y')}\,  \\
&\times&\hbox{Tr}\Bigg(
 A[I_4-D_0^{-1}A]^{-1}\arrowvert_{(x,x')}\, (\barr{q}+m_N)
n(q)A[I_4-D_0^{-1}A]^{-1}
\arrowvert_{(y',y)}(\barr{p}+m_N)\,n(p)\Bigg)+...\Bigg\}~,
\nonumber
\ea
where the trace, indicated by $\hbox{Tr}$, is over the spinor and flavor
indices.  {}From this result we can readily read off the new effective chiral
Lagrangian density $\widetilde{\La}_{\pi\pi}$ in the presence of nucleonic
densities just by equating the expression between curly brackets to $i\int dx
\widetilde{\La}_{\pi\pi}$. Furthermore, the diagonal flavor matrix $n(p)$ in
eq.\,(\ref{fZ2}) is:
\ba
 \label{as}
 n(p)&=&
 \left( \begin{array}{cc}\theta(k_F^{(p)}-|\vp|) 
 & 0\\ 0 & \theta(k_F^{(n)}-|\vp|) 
 \end{array}\right)
 \equiv \left( \begin{array}{cc}n(p)_1 & 0\\ 0 & n(p)_2 \end{array}
 \right) =\hat{n}(p)I_2+\bar{n}(p)\tau_3~,
\ea
with $I_2$ the $2\times 2$ unity matrix, $\theta(x)$ the Heaviside step
function, $\tau_3$ the usual Pauli matrix, $\tau_3 = {\rm diag}(1,-1)$,
$\hat{n}(p)=(n(p)_1+n(p)_2)/2$ and $\bar{n}(p)= (n(p)_1-n(p)_2)/2$. We are
considering the general case of asymmetric nuclear matter with two Fermi-seas
of protons and neutrons with densities $\rho^p=(k_F^{(p)})^3/3\pi^2$ and
$\rho^n=(k_F^{(n)})^3/3\pi^2$, respectively, with $k_F^{(p)}$ and $k_F^{(n)}$
the corresponding Fermi momenta. In eq.\,(\ref{fZ2}) the nucleon energy is
given by $E(p)= \sqrt{\vp^2+m_N^2}$, where $m_N=(m_n+m_p)/2\simeq 939$ MeV,
since the difference between the proton and neutron masses is $\Opd$ giving
rise to in-medium contributions of higher order than the next-to-leading order
ones, see section \ref{sec:counting}.  For the same reason, we will also use
the average nucleon mass, $m_N$, in the baryon propagators.

The ellipses in eq.\,(\ref{fZ2}) indicate terms with a higher number of
three-momentum integrals over the Fermi-seas coming from a logarithmic
expansion of the vacuum interaction operator
$\Gamma=-iA[I_4-D_0^{-1}]^{-1}$\cite{med1}. Notice that the Fermi-sea states
are {\it on-shell}, $p^0=E(p)$. Indeed, for $n\geq 1$ Fermi-sea insertions one
picks up a factor $(-1)^{n+1}/n$, while every operator
$\Gamma=-iA[I_4-D_0^{-1}]^{-1}$, containing both pion legs and external
sources, implies a phase of $-i$ . The operator $A$ is defined as follows: 
\be
 \label{defa}
 \int dx\,\La_{\pi N}=\int dx\,\bar{\psi}(x)D(x)\psi(x)=\int dx\,
 \bar{\psi}(x)D_0(x)\psi(x)-\int dx\,dy\,\bar{\psi}(x)A(x,y)\psi(y)~,
\ee
where $D_0(x)=i\gamma^\mu \partial_\mu-M_N$ is the free Dirac operator and the
symbol $M_N$ refers to the diagonal matrix $M_N={\rm diag}(m_p,m_n)$ of
physical proton and neutron masses, respectively.  Hence, standard in-medium
generalized vertices from eq.\,(\ref{fZ2}) can be represented as in
fig.\,\ref{fig:gv} where each connected diagram represents one of such
vertices and the vacuum non-local vertices $\Gamma$ are connected through the
exchange of on-shell Fermi-sea states.  The latter are represented by the
thick solid lines in that and all following diagrammatic figures. Furthermore,
$\Gamma$ results from the iteration of the $-iA$ operator with intermediate
{\it free} baryon propagators $i D_0^{-1}$ as shown in fig.\,\ref{fig:gop}.
Consequently, any diagram with medium contributions will be a set of
$V_\rho\geq 1$ generalized in-medium vertices. For each of them one has
$n_j\geq 1$ Fermi-sea insertions, $1\leq j\leq V_\rho$, $m_j\geq 0$ free
baryon-propagators and of $m_j+n_j$ vertices $-i A$.  The resulting Feynman
rules (as derived in \cite{med1}) read:

\begin{itemize}
\item First include the global sign $(-1)^{V_\rho}$ because of the fermionic
  closed loop attached to each generalized vertex, and the combinatoric sign
  factor \be \prod_{j=1}^{V_\rho}\frac{(-1)^{n_j}}{n_j} \ee from the
  logarithmic expansion giving rise to the different in-medium vertices.
\item Then, following the vertex in the opposite sense to that of the
  fermionic arrows, write for each Fermi-sea insertion an integral 
 \be
   \int\!\!\frac{d{\mathbf{p}}\, (\barr{p}+m_N)}{(2\pi)^32E(p)}n(p) 
  \ee 
  with
  $p^0=E(\mathbf{p})$ and for every vacuum baryon propagator with free
  momentum $p$ write 
  \be 
    i \int\!\!\frac{d p}{(2\pi)^4}
    \frac{\barr{p}+m_N}{p^2-m_N^2+i\epsilon}~.  
   \ee
\item For a vertex in momentum space write a term
   \be 
     -iA \, (2\pi)^4~,
    \ee
    keeping in mind the energy-momentum conservation at each vertex.
\item Then for each pion internal  line write the pion vacuum 
    propagator 
   \be
     i\int \frac{dq}{(2\pi)^4}\frac{1}{q^2-M_\pi^2+i\epsilon}~.
   \ee
\item Finally, 
    sum over the Dirac and spin indices of the fermions. 
\end{itemize}
This defines explicitly the Feynman rules in momentum space in order to obtain
$i (2\pi)^4$ times the desired connected graph accompanied by the global delta
function of energy-momentum conservation.  It is important to keep in mind
that a generalized vertex behaves as a standard {\it local} one of quantum
field theory with respect to the determination of the numerical factors
accompanying a given diagram under the exchange of pions.  This can be seen by
simply applying standard perturbative techniques in path integrals to the
action given between brackets in eq.\,(\ref{fZ2}).  It is also worthwhile to
note that one can take eq.\,(\ref{fZ2}) directly in configuration space, see
sections \ref{sec:qkcon} and \ref{sec:prop} and the appendix \ref{subsec:wfr}.


\section{Chiral Lagrangians and the interaction 
 operators $A^{(1)}$ and $A^{(2)}$}
\label{sec:operators}
\def\theequation{\Alph{section}.\arabic{equation}}
\setcounter{equation}{0}

To evaluate the next-to-leading order medium corrections to the Green
functions from the $\pi N$ Lagrangian ${\cal L}_{\pi N}=\bar{\psi}D(x) \psi$,
as discussed in detail in sec.~\ref{sec:counting}, one has to consider the
$\Op$ ${\cal L}^{(1)}_{\pi N}$ and $\Opd$ ${\cal L}^{(2)}_{\pi N}$
Lagrangians. For relativistic spin-1/2 fields chirally coupled to pions and
external sources, the lowest order effective Lagrangian reads~\cite{gass}
\ba
 \label{pin1}
 {\cal L}_{\pi N}^{(1)}=\bar{\psi}\left(i\gamma^\mu \partial_\mu
 -\krig{m}_N\! I_2+i\gamma^\mu\Gamma_\mu+i\krig{g}_A 
 \gamma^\mu \gamma_5 \Delta_\mu \right)\psi~,
\ea
with $\Delta_\mu=\frac{1}{2}u^\dagger \nabla_\mu \!U\,u^\dagger$ in terms of
the covariant derivative $\nabla_\mu U(x)=\partial_\mu U(x)-i
(v_\mu(x)+a_\mu(x))U(x)+iU(x)(v_\mu(x)-a_\mu(x))$.  $\Gamma_\mu$ is the chiral
connection, $\Gamma_\mu= \frac{1}{2}\left[u^\dagger,\partial_\mu
  u\right]-\frac{i}{2} u^\dagger
(v_\mu+a_\mu)u-\frac{i}{2}u(v_\mu-a_\mu)u^\dagger$. The Goldstone bosons (the
pions) are collected in the $2\times 2$ unitary matrix $u=\exp(i\phi/2f)$,
$U=u^2$ and $\phi$ is given by:
\ba
 \label{phi}
 \phi=\left(\begin{array}{cc}
 \pi^0 & \sqrt{2}\pi^+ \\ \sqrt{2}\pi^- & -\pi^0
 \end{array}\right).
\ea
On the other hand, the constants $\krig{m}_N$, $\krig{g}_A$ and $f$ refer to
the mass, axial coupling of the nucleon and weak pion decay constant in the
SU(2) chiral limit. Taking into account the definition of the $A$ operator,
$A(x)\equiv D_0(x)-D(x)$,\footnote{Although the operator $A$ is in general
  non-local, $A(x,y)$, we omit here the second argument since it just involves
  the delta function, $\delta(x-y)$, or derivatives thereof.} with
$D_0(x)=i\gamma^\mu \partial_\mu- M_N$, we obtain from eq.\,(\ref{pin1}) for
the first term in the chiral expansion of the operator $A$, 
\ba
 \label{a1}
 A^{(1)}=-i\gamma^\mu\Gamma_\mu-i\krig{g}_A \gamma^\mu \gamma_5 
 \Delta_\mu~.
\ea
The difference $\krig{m}_N\!I_2-M_N$ is $\Opd$ \cite{gass} and will be
included in $A^{(2)}$.  Although strictly speaking there is no explicit
symmetry breaking at leading order in the effective pion--nucleon Lagrangian,
this identification will turn out to be convenient at later stages.

To obtain the complete expression of the (second order) $A^{(2)}$ operator, we
have to take into account the Lagrangian ${\cal L}^{(2)}_{\pi N}$ (we use here
the notation of ref.~\cite{fettes}).  The operator $A^{(2)}$ can be obtained
by just removing the $\psi$ and $\bar{\psi}$ fields from the Lagrangian and by
changing the overall sign,
\ba
 \label{pin2}
 A^{(2)}&=&\krig{m}_N\!I_2-M_N -  c_1 \langle \chi_+ \rangle 
 +\frac{c_2}{2 \krig{m}_N^2}\langle u_\mu u_\nu \rangle D^\mu D^\nu
 -\frac{c_3}{2}\langle u_\mu u^\mu \rangle 
 +\frac{c_4}{4}\gamma^\mu \gamma^\nu [u_\mu,u_\nu]
 -c_5\,\hat{\chi}_+ 
\nn\\
 &-&\frac{ic_6}{8\krig{m}_N}\gamma^\mu \gamma^\nu F_{\mu\nu}^+
 - \frac{ic_7}{8\krig{m}}\gamma^\mu \gamma^\nu \langle F_{\mu\nu}^+\rangle~,
\ea

where the $\Opd$ low-energy constants $c_i$ are finite and where we have also
added the term $\krig{m}_N\!I_2-M_N$ as discussed above. Furthermore, we have
$D_\mu= \partial_\mu+\Gamma_\mu$, $u_\mu=2i\Delta_\mu$, $\chi_+=u^\dagger \chi
u^\dagger+ u\chi^\dagger u$ with $\chi(x)=2 B_0(s(x)+ip(x))$. Here, $B_0\,
\delta^{ij} = -\langle 0| \bar{q}^iq^j | 0\rangle / f^2$ measures the strength
of the symmetry breaking in the chiral limit.  The quark mass matrix ${\cal
  M}= {\rm diag}(m_u,m_d)$ is included in the scalar source $s(x)={\cal
  M}+\cdot\cdot\cdot$. The trace in flavor space of an arbitrary matrix ${\cal
  O}$ is denoted by $\langle {\cal O} \rangle$.  In this way, the traceless
matrix $\hat{\chi}_+$ is defined as $\hat{\chi}_+=\chi_+- \frac{1}{2}\langle
\chi_+ \rangle$. Finally, $F_{\mu\nu}^+=u^\dagger F_{\mu\nu}^R u + u
F_{\mu\nu}^L u^\dagger$ where $F_{\mu\nu}^L=\partial_\mu \ell_\nu-\partial_\nu
\ell_\mu-i[\ell_\mu,\ell_\nu]$, $F_{\mu\nu}^R=\partial_\mu r_\nu-\partial_\nu
r_\mu-i[r_\mu,r_\nu]$ with $\ell_\mu=v_\mu-a_\mu$ and $r_\mu=v_\mu+a_\mu$.
For more details see e.g. ref.~\cite{fettes}. We briefly collect the chiral
power (or chiral dimension) of the basic ingredients that appear repeatedly in
the chiral expansion of the operator $A$. Namely, $\partial_\mu u$, $v_\mu$
and $a_\mu$ are $\sim \Op$ while $s$, $p$, $\chi$ and $F_{\mu\nu}^{L,R}$ are
$\sim \Opd$. On the other hand, $u$ and $\partial_\mu \psi$ $\sim {\cal O}(1)$
(although $\partial_i \psi$ $\sim \Op$ for the spatial components $i$). We
point out that the first two terms in eq.\,(\ref{pin2}) appear because the
free Dirac operator $D_0$ is defined in terms of the {\it physical} nucleon
masses (and not the mass in the chiral limit).  It is straightforward to show
that up to $\Opd$ in $A^{(2)}$, the difference $\krig{m}_N\!I_2-M_N$ is
canceled by the constant contributions coming from the terms proportional to
$c_1$ and $c_5$. Up to $\Opd$, it is easy to work out from the Lagrangians
${\cal L}_{\pi N}^{(1)}$ and ${\cal L}_{\pi N}^{(2)}$ the following
expressions for the masses of protons, $m_p$, and neutrons, $m_n$: 
\ba
 m_p&=&\krig{m}_N-8B_0 c_1 \hat{m}-4 B_0 c_5 \bar{m}~,\nn \\
 m_n&=&\krig{m}_N-8B_0 c_1 \hat{m}+4 B_0 c_5 \bar{m}~, 
\ea 
where $\hat{m}=(m_u+m_d)/2$ is the average light quark mass and
$\bar{m}=(m_u-m_d)/2$ measures the strength of strong isospin breaking.  When
inserting the previous expressions in eq.\,(\ref{pin2}) the difference
$\krig{m}_N\!I_2-M_N$ will cancel those terms from $c_1$ and $c_5$ that are
independent of the quantum pion fields or the external sources $v$, $a$, $s$
and $p$. Indeed one can generalize this result for any $A^{(n)}$ operator,
with $n\geq 2$, following similar arguments to the ones given in
ref.~\cite{fettes}.  Among the building blocks for the construction of a
chiral ${\cal L}_{\bar{\psi}\psi}$ Lagrangian, only the scalar operator
$\chi_+$ gives rise to terms without pion or external source legs, leading to
a constant matrix. Given $\chi_+$, one can then build up operators ${\cal O}$
to be sandwiched between the $\bar{\psi}$ and $\psi$ fields which give rise to
constant matrices without any field. In addition, one can include any number
of covariant derivatives $D_\mu$, since they count as ${\cal O}(1)$ and the
partial derivative $\partial_\mu$, contained in $D_\mu$, does not include any
pion or external source in contrast to $\Gamma_\mu$. However, we can integrate
by parts in such a way, that all the $\partial_\mu$ act either to the left or
to the right of the $\chi_+$, as otherwise they will give rise to vertices
with at least one pion or external leg. When all the Lorentz indices of these
operators are contracted with the $\gamma-$matrices, the metric tensor and the
totally antisymmetric tensor in $d=4$ dimensions \cite{fettes}, it follows
from the application of the equations of motions and from the trivial result
$\left[\partial_\mu,\partial_\nu\right]=0$ that the $\partial_\mu$ operators
in the covariant derivatives just give rise to factors of $\krig{m}$. These
factors can be reabsorbed in the couplings of those operators that are only
constructed from $\chi_+$ or in the vertices with one or more legs. As a
result all the vertices without any legs are just constant and their finite
parts are reabsorbed in the definition of the physical nucleon mass matrix
$M_N$. The infinite parts that could arise because of possible scale dependent
counterterms in front of such terms, needed for the renormalization, have to
be considered jointly with pions loops, which obviously involve pion legs.


\section{Application of the non-standard counting rules 
to in-medium pion propagation}
\label{sec:example}
\def\theequation{\Alph{section}.\arabic{equation}}
\setcounter{equation}{0}

As an illustration we will discuss the application of the non--standard power
counting in detail, as it appears for the in-medium pion-propagation.  The
leading contribution is $\Opt$ and contains one generalized vertex, $V_\rho=1$
at lowest order, $\delta_1=4$, while $V_\pi=0$, $n_m=0$ for $m\geq 2$ and
$I_B=I_B^\star$. Obviously $p_m=0$ for all $m$ since $V_\pi=0$. This case
corresponds to figs.\,\ref{fig:pp2}a, where only one Fermi-sea line has been
shown although it should be clear that one can substitute any thin line by a
thick one including the appropriate factors as discussed in detail in
appendix~\ref{subsec:genfunc}, in sec.\,\ref{sec:prop} and in
ref.~\cite{med1}. The next-to-leading order is then $\Opc$ and there are
various contributions:

\begin{enumerate}
\item[a)] The same as before but with $\Delta I_B\equiv I_B-I_B^\star=1$. This
  corresponds to fig.\,\ref{fig:pp2}.b.  Other diagrams similar to this one
  obtained just by exchanging the positions of the lines and the loop are not
  shown. Instead of a loop one can also have an $\Opt$ $\pi N$ counterterm
  without pion lines just for renormalization, as discussed at the end of
  sec.~\ref{sec:operators}.
\item[b)] $V_\rho=1$, $\delta_1=4$, $V_\pi=0$, $n_2=1$ and $\Delta I_B=0$.
  This gives rise to the diagrams in fig.\,\ref{fig:pp2}c,d, where, as before,
  diagrams obtained under the exchange of the positions of lines and the loop
  or by the inclusion of any extra Fermi-sea insertion are not explicitly
  shown. The diagram of fig.\,\ref{fig:pp2}d is zero because the Weinberg term
  with two pions involves the vertex $\left[\phi,\partial_\mu \phi\right]=2 i
  \epsilon_{ijk}\phi_i \partial_\mu \phi_j$, and hence one cannot have the
  same two flavors in one vertex.
\item[c)]$V_\rho=2$, $\delta_{1,2}=4$, $V_\pi=0$, $n_m=0$ $m\geq 2$ and
  $\Delta I_B=0$.  The diagrams are shown in figs.\,\ref{fig:pp2}e,f. As
  before diagrams obtained by exchanging lines and adding more Fermi-sea
  insertions are not shown. The diagram in fig.\,\ref{fig:pp2}f vanishes
  because the new generalized vertex with only one pion is zero since there is
  a $\gamma_5$ matrix from the pion vertex and at most two additional gamma
  matrices.
\end{enumerate}

It is worthwhile to mention that in this example there is no contribution at
next-to-leading order from the insertion of one $A^{(2)}$ operator in a
generalized in-medium vertex, because then there are at least two pion lines
involved, so that $\delta_1=5$ {\em and} $n_2=1$. As a result such
contributions only start at $\Opf$.

In order to calculate the diagram in fig.\,\ref{fig:pp2}a one has to make use
of the full baryon propagator and not the heavy baryon (HB) expansion in
eq.\,(\ref{pop}). We have calculated this diagram for the case of symmetric
nuclear matter, where, as discussed above, the diagram fig.\,\ref{fig:pp2}c
vanishes as well. Then, up to $\Opt$ and for $\barro=0$, we have instead of
eqs.\,(\ref{spec}): 
\ba
 \label{specg}
 &&
 \omega^2-{\mathbf Q}^2-M_\pi^2
 +\frac{2m_N g_A^2 Q^2}{f^2}\left(\ci(\omega,{\mathbf Q})+
 \ci(-\omega,{\mathbf Q})\right)\nn\\
 &&
 -i\frac{g_A^2 Q^2 m_N}{2\pi f^2}\left(\theta(\omega)
 \theta(k_F^2-2m_N \omega)\theta(k_F-
 k_M(\omega))\frac{k_F^2-k^2_M(\omega)}
 {4|{\mathbf Q}|}+\omega\rightarrow -\omega\right)=0~,
\ea
where $k^2_M(\omega)$ is the maximum of $k_F^2-2 m_N \omega$ 
and $(Q^2+2m_N \omega)^2/4{\mathbf Q}^2$.  
The function $\ci(Q^0,{\mathbf Q})$ is given by:
\ba
 \label{funi}
 \ci(Q^0,{\mathbf Q})&=&{\cal P}\int^{k_F}\!\!\!\! \frac{d\mathbf{p}}{(2\pi)^3}
 \frac{1}{Q^2+2m_N Q^0-2{\mathbf p}{\mathbf Q}}\nn \\
 &=&\frac{k_F (Q^2+2m_N Q^0)}{16 \pi^2 {\mathbf Q}^2}+
 \frac{(Q^2+2m_N Q^0)^2-4k_F^2 {\mathbf Q}^2}{128\pi^2 |{\mathbf Q}|^3}\log
 \frac{(Q^2+2m_N Q^0-2k_F|{\mathbf Q}|)^2}{(Q^2+2m_N Q^0+2k_F 
 |{\mathbf Q}|)^2}~,\nn\\
\ea
where ${\cal P}$ indicates the principal value of the integral.  The step
functions in eq.\,(\ref{specg}) stem from the interplay of the second
Fermi-sea insertion in fig.\,\ref{fig:pp}c with the pole from the free baryon
propagator in fig.\,\ref{fig:pp}b as discussed in sec.~\ref{sec:prop}.


\section{Wave function renormalization}
\label{subsec:wfr}
\def\theequation{\Alph{section}.\arabic{equation}}
\setcounter{equation}{0}

The pion fields are naturally normalized by imposing the canonical 
commutation relations at equal times $x^0=y^0$:
\ba
 \label{comr}
 \left[\pi(x)^\alpha,\widetilde{\Pi}(y)^\beta\right]
 =i \delta({\mathbf{x}}-{\mathbf{y}})
 \delta^{\alpha\beta}~,\nn\\
 \left[\pi(x)^\alpha,\pi(y)^\beta\right]=\left[\widetilde{\Pi}(x)^\alpha,
 \widetilde{\Pi}(y)^\beta\right]=0~, 
\ea
where the indices $\alpha$, $\beta$ can refer to any of the $\pi^0$, $\pi^+$
and $\pi^-$ fields and $\widetilde{\Pi}(x)^\alpha$ is the conjugate momentum
of the field $\pi(x)^\alpha$.  We now proceed to calculate the in-medium pion
field and conjugate momentum from the quadratic term in the pion fields, $\int
dx \widetilde{\La}_{\phi\phi}$, of $\int dx \widetilde{\La}$.

The in-medium {\it free} pion fields are given, at the classical level, by the
equations of motion,
\[ 
 \delta \int dx \widetilde{\La}_{\phi\phi}/\delta \phi_k(y)=0~,
\] 
already discussed in sec.~\ref{sec:prop}. These are fulfilled by writing the
pion fields as:
\ba
 \label{phif}
 \pi^+(x)&=&\int\frac{d{\bf Q}}{(2\pi)^3}\left[a(\vQ)e^{-iQx}+
 b^\dagger(\vQ) e^{i\tQ x}\right]~,\nn\\
 \pi^-(x)&=&\int\frac{d{\bf Q}}{(2\pi)^3}\left[b(\vQ) e^{-i\tQ x}+
 a^\dagger(\vQ) e^{iQx}\right]~,\nn\\
 \pi^0(x)&=&\int\frac{d{\bf P}}{(2\pi)^3}\left[c(\vP) e^{-iPx}+c^\dagger(\vP) 
 e^{iPx}\right]~,
\ea   
where $Q^0(\vQ^2)=\omega_+(\vQ^2)$, $\tQ^0(\vQ^2)=\omega_-(\vQ^2)$ and
$P^0(\vP^2)=\omega_0(\vQ^2)$ are positive functions of the square of the
three-momentum and correspond, respectively, to the $\pi^+$, $\pi^-$ and
$\pi^0$ energies and are given by the corresponding spectral relation between
energy and three-momentum, eqs.\,(\ref{spec}). Note that in eqs.\,(\ref{phif})
we have space-time dependences of the form $e^{-iQx}$ and $e^{iQx}$. When
applying eqs.\,(\ref{spec}) to the latter case, we have to substitute
$Q^0\rightarrow -Q^0$ since these formulae were deduced taking a generic form
$e^{-iQx}$ and allowing for both positive as well as negative values of
$Q^0\equiv \omega$. 
Taking this into account, together with the fact that the terms
proportional to $\barro$ in eqs.\,(\ref{spec}) are odd in $\omega$, we can then
combine the coefficients $a(\vQ)$ and $b(\vQ)$ and their conjugates in the
$\pi^+$ and $\pi^-$ fields as shown in eqs.\,(\ref{phif}).  On the other hand,
the in-medium contribution to the pion conjugate momenta
$\widetilde{\Pi}_i(x)$ are calculated by differentiating the quadratic piece
$\int dx \delta \widetilde{\La}_{\phi\phi}$ (given up to $\Opf$ and times a
factor of $i$ in eq.\,(\ref{pp1})) with respect to $\dot{\phi}(x)_i$.  In
order to deal with the presence of the non-localities induced by the baryon
propagator $D_0^{-1}(x,y)$, we have worked in momentum space after calculating
the functional derivative. Proceeding in this way, one finds for
$\widetilde{\Pi}_i(x)$: 
\ba
 \label{deltapi}
 \widetilde{\Pi}(x)_i
 &=&
 \dot{\phi}(x)_i\left\{1+\frac{4\hatr}{f^2}\left( c_2+c_3-
 \frac{g_A^2}{8m_N} \right)\right\}+\sum_{k=1}^3\epsilon_{ik3}
 \frac{\barro}{2f^2}\phi_k(x) \nn \\
 &-& 
 i\frac{g_A^2\hatr}{2 m_N f^2}
 \int\frac{d{\mathbf Q}}{(2\pi)^3}\,
 \frac{{\mathbf Q}^2}{2 \omega_{{\rm vac}}(\vQ^2)^2}\left[
 \phi_i(\vQ) e^{-iQx}\!-\!\phi_i(\vQ)^\dagger e^{i Qx}\right]~,
\ea
where we have denoted the vacuum energy by 
$\omega_{{\rm vac}}(\vQ^2)=\sqrt{M_\pi^2+\vQ^2}$.
One can easily calculate the conjugate momenta of the fields $\pi^+(x)$,
$\pi^-(x)$ and $\pi^0(x)$ from eq.\,(\ref{deltapi}). Taking also into account
eqs.\,(\ref{phif}) one has:
\ba
 \label{cmcharged}
 \widetilde{\Pi}(x)^+&\!\!\!=\!\!\!&-i\int\frac{d\vQ}{(2\pi)^3}
 \left[\omega_-(\vQ^2) b(\vQ) 
 e^{-i\tQ x}
 -\omega_+(\vQ^2)a^\dagger(\vQ)e^{iQx}\right]\left\{1+\frac{4\hatr}{f^2}
 \left( c_2+c_3-
 \frac{g_A^2}{8m_N}\frac{M_\pi^2}
 {\omega_{{\rm vac}}(\vQ^2)^2} \right)\right\} \nn\\
&-&
 i\frac{\barro}{2f^2}\pi^-(x)~,\nn\\
 \widetilde{\Pi}(x)^-&\!\!\!=\!\!\!&-i\int\frac{d\vQ}{(2\pi)^3}
 \left[\omega_+(\vQ^2) a(\vQ) 
 e^{-i Q x}-\omega_-(\vQ^2)b^\dagger(\vQ)e^{i\tQ x}\right]
 \left\{1+\frac{4\hatr}{f^2}
 \left( c_2+c_3-
 \frac{g_A^2}{8m_N}\frac{M_\pi^2}
 {\omega_{{\rm vac}}(\vQ^2)^2} \right)\right\} \nn\\
&+&
 i\frac{\barro}{2f^2}\pi^+(x)~,\nn\\
 \widetilde{\Pi}(x)^0&=&-i\int\frac{d\vP}{(2\pi)^3}\omega_0(\vP^2)\left[c(\vP) 
 e^{-i P x}-c^\dagger(\vP)e^{i P x}\right]\left\{1+\frac{4\hatr}{f^2}
 \left( c_2+c_3-
 \frac{g_A^2}{8m_N}\frac{M_\pi^2}
 {\omega_{{\rm vac}}(\vP^2)^2} \right)\right\}~. \nn\\
\ea

We now proceed to the quantization of the pion fields eqs.\,(\ref{phif}) by
imposing the canonical commutation relations given in eqs.\,(\ref{comr}) for
equal times. We denote by $d_\alpha(\vQ)$ a generic coefficient in
eqs.\,(\ref{phif}) and by $d_\alpha^\dagger(\vQ)$ its complex conjugate, with
$\alpha=0$ referring to a $\pi^0$ and $\alpha=+(-)$ to a $\pi^+$ ($\pi^-$)
state. We quantize them by imposing the following commutations relations:
\ba
 \label{ans}
 \left[d_\alpha(\vQ),d_\beta(\vP)\right]
 =(2\pi)^3 \,N_\alpha(\vQ^2) \delta(\vQ-\vP) 
 \delta_{\alpha \beta}~,\nn\\
 \left[d_\alpha(\vQ),d_\beta(\vP)\right]
 =\left[d^\dagger_\alpha(\vQ),d^\dagger_\beta (\vP)\right]=0~.
\ea 
We now prove that it is possible to satisfy eqs.\,(\ref{comr}) by an
appropriate choice of $N_\alpha(\vQ^2)$.

$N_0(\vQ^2)$ is fixed by the conditions $x^0=y^0$ and
$\left[\pi^0(x),\widetilde \Pi^0(y)\right]=i \delta(\vx-\vy)$ with the result:
\be
 \label{n3}
 N_0(\vQ^2)=\frac{1}{2 \omega_0 Z(\vQ^2)}~,
\ee
where 
\be
 \label{z0}
 Z(\vQ^2)=1+\frac{4\hatr}{f^2}\left( c_2+c_3-
   \frac{g_A^2}{8m_N}\frac{M_\pi^2}{\omega_{{\rm vac}}(\vQ^2)^2} \right).  
\ee 
The other commutation relations involving the $\pi^0(x)$ fields are then
trivially satisfied. In the case of the charged pion fields, the commutation
relations at equal times read: $\left[\pi^+(x),\pi^-(y)\right]=0$ and
$\left[\pi^+(x),
  \widetilde{\Pi}^+(y)\right]=\left[\pi^-(x),\widetilde{\Pi}^-(y)\right]=i
\delta(\vx-\vy)$. These conditions imply the following equations:
\ba
 &&N_+(\vQ^2)=N_-(\vQ^2)~,\nn\\
 &&\omega_+(\vQ^2)N_+(\vQ^2)+\omega_-(\vQ^2)N_-(\vQ^2)=Z(\vQ^2)^{-1}~.
\ea
Thus:
\be
 N_+(\vQ^2)=N_-(\vQ^2)=\frac{1}{[\omega_+(\vQ^2)+\omega_-(\vQ^2)]Z(\vQ^2)}~.
\ee

We are now in the position to discuss the wave function renormalization. We
will normalize our pion states with definite three-momentum $\vp$,
$|\pi^\alpha(\vp)\rangle$, as:
\be
 \label{nor}
 \langle\pi^\alpha(\vq)|\pi^\alpha(\vp)\rangle
 =2\omega_\alpha(\vq^2)(2\pi)^3\delta(\vp-\vq)~.
\ee
Here we have chosen the functions in front of $\delta(\vp-\vq)$ as in the
vacuum, so that the calculated matrix elements do not change with respect to
the vacuum ones in the limit $k_F\rightarrow 0$. On the other hand, we define
the one-pion state in terms of the creation operators as:
\be
 \label{pstates}
 |\pi^\alpha(\vp)\rangle
 =\Theta_\alpha(\vp^2) d^\dagger_\alpha(\vp)|\Omega\rangle~. 
\ee
Imposing eq.\,(\ref{nor}), together with eqs.\,(\ref{ans}), one has (except
for a multiplicative phase that can always be chosen to be one):
\be
 \label{tetas}
 \Theta_\alpha(\vp^2)=\sqrt{\frac{2\omega_\alpha(\vp^2)}{N_\alpha(\vp^2)}}~.
\ee
In the calculation of a matrix element of a certain process with asymptotic
pions, the latter are annihilated or created by the corresponding pion fields.
In this way, taking into account eqs.\,(\ref{phif}), (\ref{ans}),
(\ref{pstates}) and (\ref{tetas}), one finds that:
\be
 \label{wfr}
 \langle \Omega|\pi^\alpha(x)|\pi^\beta\rangle
 =\sqrt{2\omega_\alpha N_\alpha(\vp^2)} 
 \,\delta_{\alpha\beta}\, 
 e^{-ipx}\equiv Z_\alpha(\vp^2)^{-\frac{1}{2}}\, \delta_{\alpha\beta}
 \,e^{-ipx}~,
\ee  
with $Z_\alpha(\vp^2)$ the wave function renormalization of the pion states of
type $\alpha$ in the nuclear medium. As a result we can proceed directly by
annihilating and creating asymptotic pion fields from the fields
$\pi^\alpha(x)$, neglecting the factors $Z_\alpha^{-\frac{1}{2}}$ during the
calculation, but multiplying in the end by as many of such factors as there
are asymptotic pion states, 
\be
 \label{wfr2}
 \prod_\alpha Z_\alpha(\vp^2)^{-\frac{1}{2}}.
\ee
The explicit expressions for $Z_\alpha(\vp^2)$ are:
\ba
 \label{z}
 Z_+(\vp^2)
 &=&
 1+\frac{4\hatr}{f^2}\left(c_2+c_3-\frac{g_A^2}{8m_N}
 \frac{M_\pi^2}{\omega_{{\rm vac}}(\vp^2)^2}\right)
 -\frac{\barro}{2f^2\omega_{{\rm vac}}\vp^2)}
 +\frac{g_A^2 \barro\, \vp^2}{2f^2\omega_{{\rm vac}}(\vp^2)^3}~,
 \nn\\
 Z_-(\vp^2)
 &=&
 1+\frac{4\hatr}{f^2}\left(c_2+c_3-\frac{g_A^2}{8m_N}
 \frac{M_\pi^2}{\omega_{{\rm vac}}(\vp^2)^2}\right)
 +\frac{\barro}{2f^2\omega_{{\rm vac}}(\vp^2)}
 -\frac{g_A^2 \barro\, \vp^2}{2f^2\omega_{{\rm vac}}(\vp^2)^3}~,
\nn\\
 Z_0(\vp^2)
 &=&
 1+\frac{4\hatr}{f^2}
 \left(c_2+c_3-\frac{g_A^2}{8m_N}\frac{M_\pi^2}
 {\omega_{{\rm vac}}(\vp^2)^2}\right)~.
\ea
It is worthwhile to point out that the wave function renormalization in
eqs.\,(\ref{z}) cannot be obtained from eqs.\,(\ref{spec}) just by calculating
their derivatives with respect to $\omega^2$ at a value determined by the
equations of motion (for a given three momentum).  The reason is the non-local
character of $\delta \widetilde{\La}_{\phi\phi}$, see eq.\,(\ref{pp1}). This
implies the presence of extra factors of energy, $\omega$, that stem from the
propagators of the nucleons and have nothing to do with time derivatives of
the pion fields, as in standard local quantum field theory.

We do not discuss the vacuum contribution to the wave function
renormalization, since we take directly the final $\Opc$ results from
\cite{gl} and since the effects of mixing the vacuum and in-medium wave
function renormalizations are of higher orders, $\Ops$ in the standard
counting.  In the latter reference the vacuum Green functions of the external
sources were directly calculated in terms of the generating functional up to
one pion loop. One has to stress that the calculation of the Green functions
of the external sources connects directly with real experiments where the
nuclear medium is tested by means of external probes, e.g.\ muon capture,
neutrino scattering, etc. This is so because the vector source $v$ can be
related to a photon and the axial-vector $a$ to the $W^{\pm}$ gauge bosons of
the weak interactions. In this way one can incorporate electromagnetic and
semileptonic weak interactions in the formalism, see e.g. ref.~\cite{pich}.

\end{appendix}

\end{document}